\documentclass[preprint,12pt,sort&compress]{elsarticle}

\usepackage{etoolbox}

\makeatletter
\patchcmd{\pprintMaketitle}
  {\footnotesize\itshape\elsaddress\par\vskip36pt}
  {\footnotesize\itshape\elsaddress\par\vskip16pt}
  {}{}

\patchcmd{\pprintMaketitle}
  {\hrule\vskip12pt}
  {\hrule\vskip6pt}
  {}{}
\makeatother

\usepackage{amssymb}
\usepackage{amsmath}

\usepackage{lineno}

\usepackage{booktabs}
\usepackage{makecell}
\usepackage{multirow} 
\usepackage{caption}
\usepackage{placeins}   
\usepackage[normalem]{ulem} 
\usepackage{hyperref}
\usepackage{xurl}
\usepackage{xcolor}   
\newcommand{\revised}[1]{\textcolor{black}{#1}}

\newenvironment{revisedblock}{\color{black}}{}

\journal{Mechanical Systems and Signal Processing}

\begin{document}

\begin{frontmatter}




\title{A unified framework for equation discovery and dynamic prediction of hysteretic systems}

\author[1]{Siyuan Yang}
\ead{syang724@connect.hkust-gz.edu.cn}

\author[1,2]{Wei Liu}
\ead{weiliu@u.nus.edu}

\author[1,3]{Zhilu Lai\corref{cor}}
\ead{zhilulai@ust.hk}

\cortext[cor]{Corresponding author.}

\affiliation[1]{organization={Internet of Things Thrust, The Hong Kong University of Science and Technology (Guangzhou)},
            country={China}
            }

\affiliation[2]{organization={Department of Industrial Systems Engineering and Management, National University of Singapore},
            country={Singapore}
            }

\affiliation[3]{organization={Department of Civil and Environmental Engineering, The Hong Kong University of Science and Technology},
            country={China}
            }



\begin{abstract}
Hysteresis is a nonlinear phenomenon with memory effects, where a system's output depends on both its current state and past states. It is prevalent in various physical and mechanical systems, such as yielding structures under seismic excitation, ferromagnetic materials, and piezoelectric actuators. 
Analytical models like the Bouc-Wen model are often employed but rely on idealized assumptions and careful parameter calibration, limiting their applicability to diverse or mechanism-unknown behaviors. 
Existing equation discovery approaches for hysteresis are often system-specific or rely on predefined model libraries, which limit their flexibility and ability to capture the hidden mechanisms. 
{To address these challenges, this research classifies equation discovery problems for hysteretic systems and develops a unified framework in which the state-space form is reformulated, and hysteretic variables are treated as trainable parameters from data. The framework further employs symbolic regression (SR) to automatically recover explicit governing equations without relying on predefined libraries, unlike methods such as sparse identification of nonlinear dynamics (SINDy). Experimental results demonstrate that the proposed method is effective in recovering governing equations for hysteretic systems, even in a challenging \textit{Full Equation Discovery} setting, where prior information is extremely limited, and solving the equations naturally enables the dynamic prediction of hysteretic systems.}
\end{abstract}





\end{frontmatter}



\section{Introduction}
Hysteresis is a common nonlinear phenomenon observed in a broad spectrum of engineering and physical systems, including ferromagnetic materials~\cite{jiles1986theory}, piezoelectric actuators~\cite{al2008modeling},
structural materials~\cite{bertotti1998hysteresis}, damping devices~\cite{ko1999design,dominguez2004modelling}, and shape memory alloys~\cite{ortin2002hysteresis}. 
Hysteresis, characterized by its path-dependent behavior, refers to systems in which the output depends not only on the current state but also its history, often appearing as input--output loops associated with energy
dissipation and irreversible processes~\cite{wen1976method,visone2008hysteresis}.
In nonlinear dynamic systems, hysteresis presents significant challenges for modeling, control, and prediction, owing to its multivalued behavior, rate dependence, and sensitivity to input history. Accurately capturing
hysteresis behavior is therefore crucial in applications such as structural health monitoring~\cite{chatzi2010experimental}, smart material design~\cite{chandra2023discovery}, and precision
actuation~\cite{liu2022experimental,vlachas2021two}.

In response to these needs, numerous mathematical models have been proposed over the past decades to characterize hysteretic behaviors. 
Among these, the Bouc-Wen model~\cite{wen1976method} has been widely used as a baseline in structural dynamics due to its compact differential representation and the ability of its parameter set to reproduce a range of
smooth, rate-dependent hysteresis loops. Its parameters provide physical interpretability and offer flexibility in capturing both softening and hardening behaviors. 
{Recent reviews and comparative works have emphasized that the classical Bouc-Wen formulation has well-known limitations when modeling more complex phenomena~\cite{capuano2022phenomenological}, including pronounced
asymmetry, pinching, and strength/stiffness degradation.
To address these, generalized Bouc-Wen models~\cite{capuano2024generalized}, and alternative phenomenological frameworks~\cite{vaiana2023analytical} have been developed and systematically compared in the recent
literature~\cite{vaiana2023classification}.
In this research, the Bouc-Wen model is used primarily as a representative benchmark to motivate the subsequent equation discovery setting.
These limitations motivate the need for more flexible, data-driven approaches that can infer governing equations directly from observed hysteretic responses.
}

In recent years, data-driven modeling has opened new venues for discovering governing laws directly from measured
data~\cite{rudy2017data,CHEN2025118330,haywood2024discussing,YI2025112149,ijcai2025p879,yan2025deep,yu2025grammar}, \revised{especially when governing mechanism of the system is unknown.} 
The sparse identification of nonlinear dynamics (SINDy)~\cite{brunton2016discovering} has shown strong potential in extracting compact and interpretable dynamic equations, by applying sparse regression with a predefined
library of candidate functions. 
Extensions such as SINDy with control~\cite{brunton2016discovering}, physics-informed SINDy~\cite{champion2019discovery,chen2021physics,liu2023interpretable}, and a sparse structural system identification method applied to
hysteretic systems~\cite{lai2019sparse} have significantly expanded its applicability to control systems, hidden-variable models, constrained physical systems, and hysteretic dynamics (our focus in this paper). 
Nevertheless, the effectiveness of SINDy-type methods strongly relies on the expressiveness of the chosen function library, which constrains their ability to capture non-polynomial dynamics, rate-dependent behaviors, and discontinuities. 
Selecting an appropriate function library is itself a nontrivial task, often requiring prior knowledge or trial-and-error tuning. 

\revised{Even in extended variant methods such as weak SINDy~\cite{messenger2021weak} and implicit SINDy~\cite{kaheman2020sindy}, sparse regression still primarily aims to identify governing equations from a prescribed candidate
library, rather than directly reconstructing hysteretic trajectories. In hysteretic systems, many essential unobserved states are involved; in this paper, these are referred to as \textit{hysteretic variables}. Unlike
auxiliary hidden quantities in standard sparse regression settings, these variables are memory-dependent states that must be inferred consistently over time from the observed response to capture the system's implicit
evolution. This makes the identification problem fundamentally more challenging than standard hidden-state or implicit-form sparse regression, and consequently limits the suitability of these methods for hysteretic
systems.}

\begin{figure} [!t]
    \centering
    \includegraphics[width=1.0\linewidth]{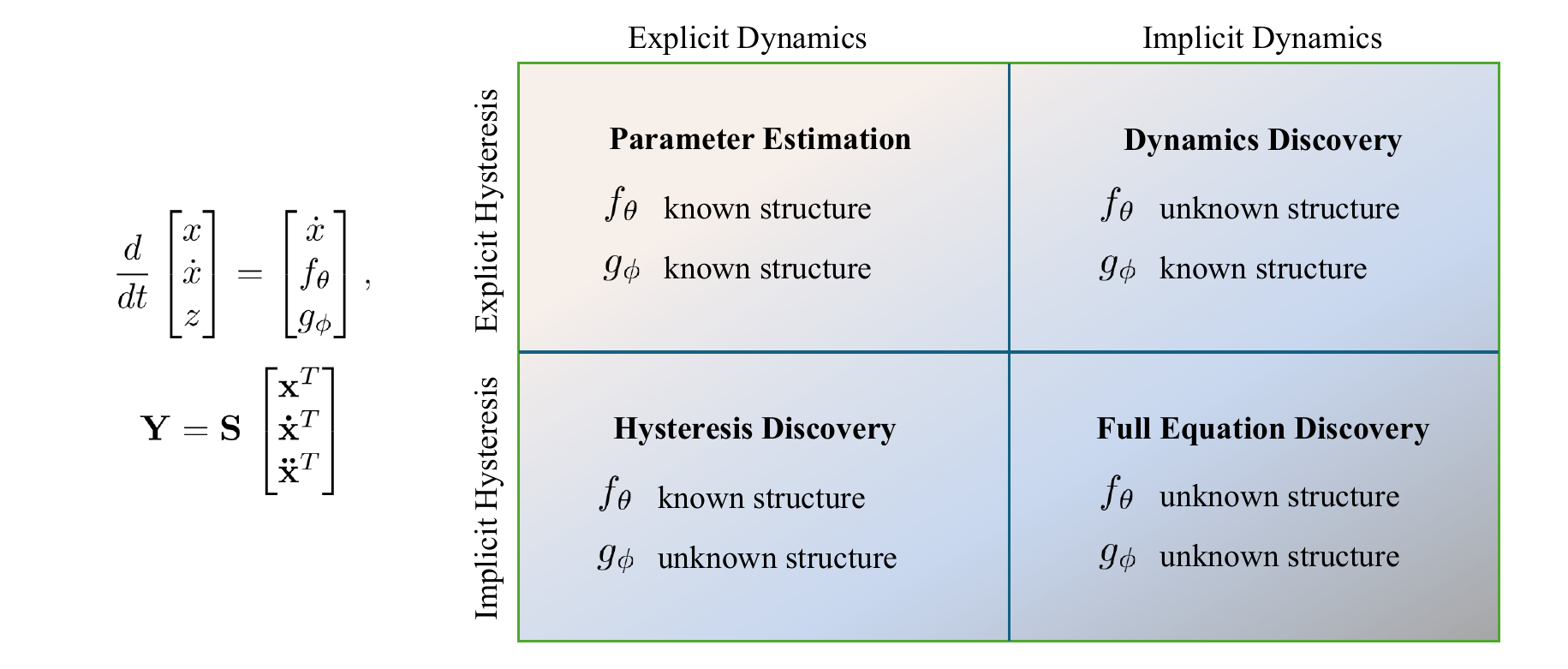}
    \caption{Classification of equation discovery approaches for hysteretic systems according to the assumed knowledge of the primary dynamics motion equation ($f_\theta$) and the hysteretic link equation ($g_\phi$). 
    \revised{The displacement variable is represented by $\mathbf{x}=[x(t_1),x(t_2),\ldots,x(t_N)]^T$, where $N$ denotes the total number of sampled time points. The velocity variable $\dot{\mathbf{x}}$ and acceleration variable $\ddot{\mathbf{x}}$ are defined analogously. 
    The selection matrix $\mathbf{S}\in\mathbb{R}^{j \times 3}$ specifies which quantities are observed, where $j=1,2\text{ or }3$ is the number of available response components among displacement, velocity, and acceleration. 
    Accordingly, $\mathbf{Y}\in\mathbb{R}^{j \times N}$ denotes the corresponding available measurements.}
    All these variables are dependent on $t$, which is omitted here for brevity. Each quadrant represents a level of prior structural specification.}
    \label{fig:structure}
\end{figure}

A variety of approaches have been proposed for equation discovery in hysteretic systems, which in this paper are classified according to the extent of prior knowledge assumed about the governing equations.  
In  Figure \ref{fig:structure}, we organize the literature into four quadrants based on whether the structure of the primary dynamics equation ($f_\theta$) and the hysteretic link equation ($g_\phi$) are explicitly known or
unknown. In the majority existing studies~\cite{lai2019sparse,bolourchi2015studies}, structural dynamics is assumed to follow an explicit 
known forms, while the hysteretic component is represented by an additional differential equation whose functional forms are to be discovered.  
Under this setting, the equation discovery task is reduced to symbolically identifying the hysteretic link equation (\textit{Hysteresis Discovery }quadrant)~\cite{wang2025symbolic,liu2025structured} or estimating
parameters when structures of both equations are known (\textit{Parameter Estimation} quadrant)~\cite{lin2022identification,worden2012parameter,charalampakis2008identification}.  
Such assumptions greatly simplify the problem but restrict applicability to systems where the underlying structural model is already well characterized.
On the other hand, frameworks that assume unknown structures in the primary dynamics equation (\textit{Dynamics Discovery} quadrant) are far less common, particularly when hysteretic effects are present.  
When both the structural dynamics and the hysteretic link equations are unknown (\textit{Full Equation Discovery} quadrant), the problem becomes substantially more challenging, requiring the simultaneous inference of two coupled, potentially nonlinear and memory-dependent governing equations. 

\revised{It is noted that the above \textit{Full Equation Discovery} setting is inherently less constrained than cases where part of the governing structure is known. The recovered symbolic forms and coefficients are commonly influenced by the operating regime and excitation levels represented in the training data, and identifiability is not fully guaranteed without sufficiently informative inputs and adequate sensing. 
Consequently, the physical interpretability of the explicitly recovered symbolic functional forms and coefficients, generalization to unseen excitation types or amplitude/frequency ranges, as well as robustness to measurement noise and modest parameter drifts, must be assessed rather than assumed. 
To the best of our knowledge, existing hysteresis equation discovery studies typically rely on partial structural assumptions (e.g.,~known primary dynamics or hysteretic link), and fully symbolic recovery of both the primary dynamics and hysteretic link in a coupled setting remains underexplored.}

\revised{The above analysis highlights two major challenges in hysteretic systems: existing methods struggle to learn unobservable hysteretic variables when the hysteretic structure is entirely unknown, and library-based methods are limited in discovering governing equations with complex nonlinear expressions.}
In this research, we propose a unified framework that integrates hysteretic variable learning and symbolic regression (SR) to address the challenges of hysteretic equation discovery. The framework enables direct inference of hysteretic variables and recovery of explicit and interpretable governing equations without relying on predefined model libraries. Our contributions are as follows: 
(i) We provide a classification of equation discovery approaches for hysteretic systems and establish a unified framework that accommodates different levels of prior knowledge;
\revised{
(ii) We introduce a solver-based hysteretic variable learning scheme, where the state-space system is reformulated to enable inference of the trainable hysteretic variable, which is not directly measurable, and then combine it with SR to recover explicit governing equations from data without relying on a fixed predefined library;
}
(iii) We validate the framework on synthetic benchmarks, a complex structure with high-order nonlinear and fractional terms, shake-table experiments of a nonlinear structure, \revised{and a multiple degree-of-freedom (MDOF) system}, demonstrating superior accuracy and generalizability compared with existing methods.

The remainder of this paper is organized as follows. Section \ref{sec2} provides necessary preliminaries. Section \ref{sec3} introduces the proposed methodology in detail. Section \ref{sec4} presents numerical and experimental validation results. Section \ref{sec5} concludes the paper with a summary of findings and directions for future research.

\section{Preliminaries} \label{sec2}
\subsection{Hysteretic models}
In modeling hysteretic systems, hysteretic variables are commonly introduced to mathematically represent memory effects, history-dependent behavior, which allows the system's response to account for past states.

Among hysteresis models, the Bouc-Wen model~\cite{wen1976method} is one of the most widely used, owing to its simplicity, and its capacity to represent a broad range of hysteretic responses. It provides a set of
differential equations to describe the evolution of hysteretic variables, and is capable of capturing both rate-dependent and asymmetric behaviors. The dynamics motion equation for a single degree-of-freedom (SDOF)
nonlinear mass--spring-damper system with hysteresis can be expressed as follows:
\begin{equation}
\label{eq:motion}
m \ddot{x}(t) + c \dot{x}(t) + F(t) = u(t),
\end{equation}
where $x(t)$ is the displacement response; $u(t)$ is the external excitation; $m$ is the mass; $c$ is the damping coefficient; and $F(t)$ is the nonlinear restoring force that accounts for both elastic and hysteretic contributions. The restoring force is defined as:
\begin{equation}
\label{eq:F}
F(t) = k x(t) + \alpha z(t),
\end{equation}
where $k$ is the stiffness, $\alpha$ is the hysteretic coefficient, and $z(t)$ is the hysteretic variable. 
Substituting 
Eq.~\eqref{eq:F} into Eq.~\eqref{eq:motion}, the equation of motion is rewritten as:
\begin{equation}
\label{exp: eq_motion}
m \ddot{x}(t) + c \dot{x}(t) + k x(t) + \alpha z(t) = u(t).
\end{equation}
Notably, setting $\alpha = 0$ reduces the equation to classical linear behaviors.

The evolution of the hysteretic variable $z(t)$ is governed by the Bouc-Wen hysteresis differential equation, which can also be written as an explicit function of $\dot{x}(t)$ and $z(t)$ for compactness and interpretability (for conciseness, we drop $t$ in the equation):
\begin{equation}
\label{exp: hysteretic_link}
\dot{z}(\dot{x}, z) = A\dot{x} - \beta|\dot{x}||z|^{n-1}z - \gamma\dot{x}|z|^n.
\end{equation}

In this formulation, the parameter $A$ serves as a linear amplification factor that scales the influence of the displacement rate on the evolution of the hysteretic variable. 
The coefficients $\beta$ and $\gamma$ jointly shape the energy dissipation and curvature of the hysteresis loop, with $\beta$ primarily influencing the width and dissipation of the loop, while $\gamma$ controls its nonlinear curvature. 
In most existing research~\cite{qian2024discovering,lai2019sparse,liu2025structured}, the exponent $n$ is typically restricted to integer values. 
\revised{In practical Bouc-Wen modeling, however, allowing non-integer values of $n$ provides a more flexible phenomenological parametrization for representing hysteretic loops with diverse transition smoothness and curvature that may not be adequately captured by integer exponents alone.}
In this research, we generalize the formulation by allowing $n$ to take more general, including fractional values, thereby enabling richer characterization of various hysteresis.


\subsection{Solvers of dynamic system}
The governing equations of nonlinear dynamic systems are rarely solvable in closed form, and thus numerical integration schemes are commonly employed to approximate their time evolution. Classical approaches include
implicit and explicit methods such as the Newmark-$\beta$ method, widely used in structural dynamics~\cite{newmark1959method}, linear multistep methods including Adams--Bashforth and Adams--Moulton
schemes~\cite{butcher2016numerical}, and the family of Runge--Kutta algorithms~\cite{iserles2009first}. Each solver provides a trade-off between accuracy, stability, and computational efficiency, and the choice of method
is usually problem dependent.   

For hysteretic models such as the Bouc-Wen formulation, the dynamics are described by a set of coupled nonlinear ordinary differential equations (ODEs) involving both the displacement response and an hysteretic variable. These equations can be integrated using standard time-stepping schemes, provided that sufficient resolution is maintained to capture the nonlinear hysteretic transition. In practice, explicit solvers are often favored for their simplicity and efficiency when the system stiffness is moderate. 

Among these available options, the classical fourth-order Runge--Kutta (RK4) method is particularly attractive, offering a good balance between accuracy and computational cost. It is easy to implement and has been widely applied in physics and engineering for nonlinear system simulations. 
In this research, the RK4 scheme is embedded into the hysteretic variable learning process to integrate the hysteretic dynamics, yielding reliable reference trajectories that facilitate subsequent equation discovery.

\subsection{Symbolic regression}
Symbolic regression (SR) is a data-driven modeling approach that seeks to uncover the underlying mathematical relationships governing a system by simultaneously identifying both the model structure and its
parameters~\cite{cranmer2023interpretable}. Unlike conventional regression methods, SR does not assume a predefined functional form and focuses solely on parameter estimation. SR performs a global search over the space of
possible mathematical expressions -- typically represented as expression trees -- to find the model that best fits the data. This search is guided by principles of accuracy, parsimony, and interpretability.

Mathematically, given a set of input--output data pairs $\{(\mathbf{x}_i, y_i)\}_{i=1}^{N} $, SR seeks to find an analytical function $h(\cdot) \in \mathcal{H}$ such that:
\begin{equation}
y_i \approx h(\mathbf{x}_i; \revised{\theta}), \quad \forall i = 1, \dots, N,
\end{equation}
where $\mathcal{H}$ denotes the space of candidate symbolic expressions constructed from a predefined set of mathematical operators (e.g.,~$+, -, \times, \div, \sin, \cos ..$.), and $\revised{\theta}$ represents any tunable parameters involved in the expression. The optimal expression $h^{*}$ is typically obtained by solving:
\begin{equation}
h^{*} = \arg\min_{h \in \mathcal{H}} \left(\mathcal{L}(h) + \lambda \, \Omega(h)\right),
\end{equation}
where $\mathcal{L}(h)$ is a loss function measuring the prediction error (e.g.,~mean squared error); $\Omega(h)$ is a complexity penalty term (e.g.,~number of nodes in the expression tree); and $\lambda$ is a regularization coefficient controlling the trade-off between accuracy and simplicity.

A feature of SR is its ability to discover explicit closed-form expressions without requiring prior specification of the model structure~\cite{quade2016prediction}. This property is particularly desirable in scientific
discovery tasks where transparency and analytical interpretability are truly essential. Unlike black-box models such as neural networks, SR yields interpretable equations that can be directly analyzed, validated, or
modified by domain experts. 
Recent algorithmic advances, including genetic programming, Monte Carlo tree search, and hybrid gradient-evolution strategies, have substantially improved both the computational efficiency and the overall expression quality of SR-generated models.
Among these, notably, PySR~\cite{cranmer2023interpretable} integrates evolutionary search with just-in-time compiled operators, thereby enabling scalable and efficient exploration of large expression spaces in scientific
applications.

\section{Methodology}  \label{sec3}
\begin{figure} [!htbp]
    \vspace*{-2.5cm} 
    \centering
    \includegraphics[width=0.96\linewidth]{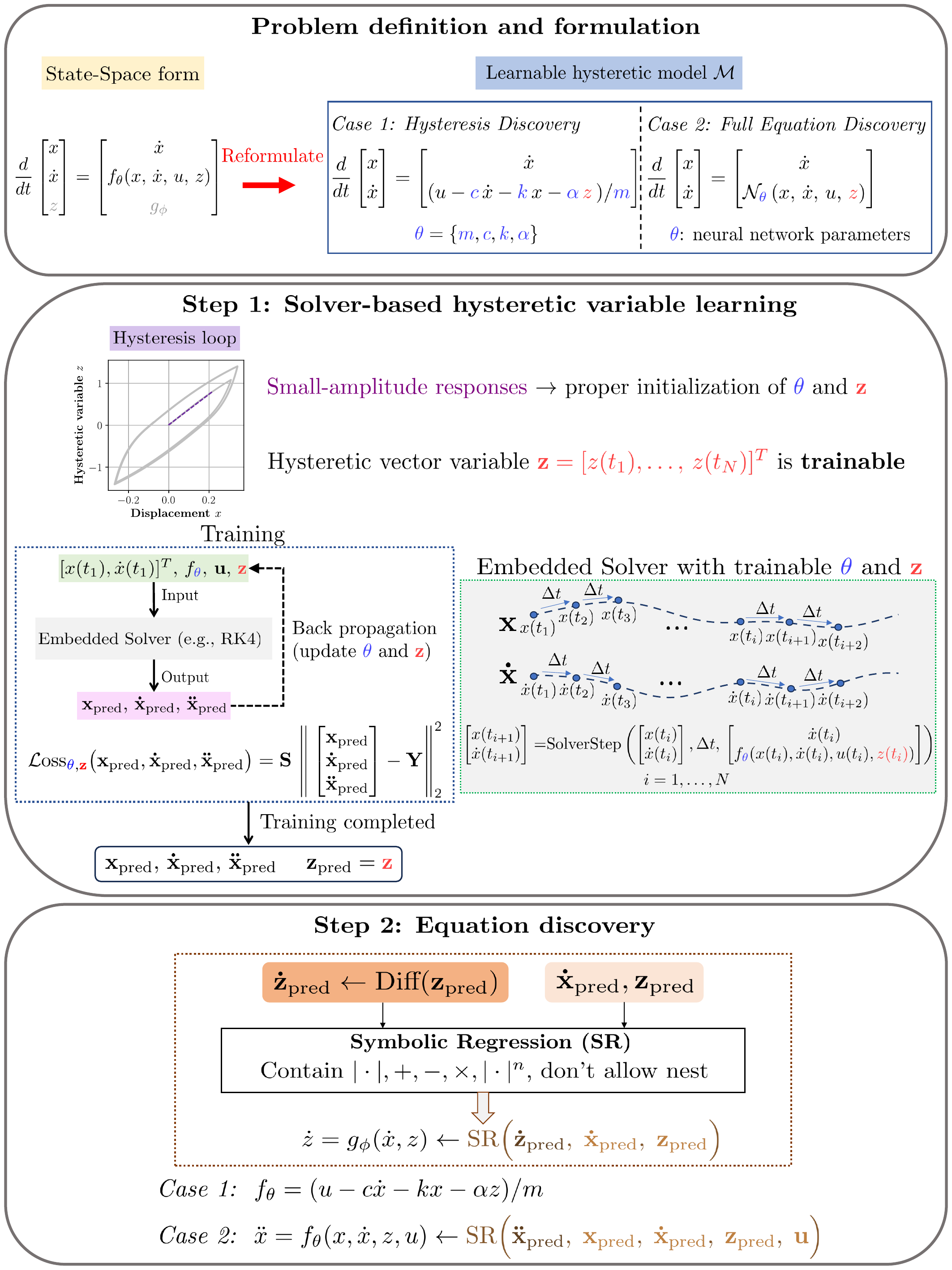}
    \caption{The proposed unified framework for equation discovery and dynamic prediction of hysteretic systems.  
    \textbf{Problem definition and formulation:} \revised{The reformulated state-space form and learnable hysteretic model are established to describe the dynamic systems.} Two cases in  Figure \ref{fig:structure} are considered as typical examples, where the hysteretic link equation ($g_\phi$) is unknown. 
    \textbf{Step 1:} \revised{The hysteretic vector variable $\mathbf{z}$ is treated as trainable parameter.} Small-amplitude responses are employed to \revised{proper initialization}. \revised{A solver is embedded to update $\theta$ and $\mathbf{z}$ when back propagation,} with the selection matrix $\mathbf{S}$ (as in Figure \ref{fig:structure}) specifying the observed components in the loss function.
    \textbf{Step 2:} Symbolic regression (SR) extracts explicit and interpretable governing equations for both the dynamics motion equation ($f_\theta$) (if \textit{Case 2}) and hysteretic link equation ($g_\phi$).}
    \label{fig:framework}
\end{figure}

\begin{table}[!htbp]
    \centering
    \captionsetup{labelfont={color=black}, textfont={color=black}}
    {\revised{%
    \caption{Problem definition. Two cases of equation discovery approaches in Figure \ref{fig:structure} for hysteretic systems. $f_\theta$: Dynamics motion equation. $g_\phi$: Hysteretic link equation.}
    \resizebox{1.0\textwidth}{!}{%
    \begin{tabular}{ccc}
        \toprule
        Type & {\makecell[c]{\textit{Case 1:}\\\textit{Hysteresis Discovery}}} & {\makecell[c]{\textit{Case 2:}\\\textit{Full Equation Discovery}}} \\
        \midrule
        Structure of $f_\theta$     & {\makecell[c]{Known (with unknown $m,c,k,\alpha$)}}    & Unknown   \\[3pt]
        Structure of $g_\phi$   & \multicolumn{2}{c}{Unknown}   \\[3pt]
        Expression of $f_\theta$   & {\makecell[c]{$f_\theta = (u - c\dot{x} - kx - \alpha z)/m$}}   & {\makecell[c]{$f_\theta = \mathcal{N}_\theta(x, \dot{x}, u, z)$}}   \\[3pt]
        Expression of $\mathbf{z}$  & \multicolumn{2}{c}{Trainable vector variable} \\[3pt]
        Initialization of $m,c,k,\alpha$  & \multicolumn{2}{c}{{\makecell[c]{Both \textit{Case 1} and \textit{Case 2} are required,\\estimated by $m\ddot{x}(t) + c\dot{x}(t) + kx(t) = u(t)$ using\\least-squares method, $\alpha$ is randomly initialized}}}  \\[3pt]
        Initialization of $\theta$   & {\makecell[c]{$\theta=\{m,c,k,\alpha\}$,\\initialized like above}}     & {\makecell[c]{$\theta$: neural network parameters,\\randomly initialized }}   \\[3pt]
        Initialization of $\mathbf{z}$  & \multicolumn{2}{c}{$z(t) = [u(t) - m\ddot{x}(t) - c\dot{x}(t) - kx(t)]/\alpha$}   \\
        \bottomrule
    \end{tabular}
}
\label{tab:cases}
}}
\end{table}

\subsection{Problem definition and formulation}
In this research, our objective is to output explicit differential equation representations of hysteretic systems, with both equation discovery and dynamic prediction.  
The proposed framework first introduces a state--space formulation:
\begin{revisedblock}
\begin{equation}
\frac{d}{dt}
\begin{bmatrix}
x \\
\dot{x} \\
z
\end{bmatrix}
=
\begin{bmatrix}
\dot{x} \\
f_{\theta}(x,\,\dot{x},\,u,\,z) \\
g_{\phi}
\end{bmatrix}
,
\end{equation}
where $x$ denotes the displacement, and $\dot{x}$ is the velocity.
\end{revisedblock}
$f_\theta$ denotes the dynamic motion equation and $g_\phi$ represents the hysteretic link equation. 
\revised{Our goal is to identify the explicit expressions of both $f_\theta$ and $g_\phi$.}
The proposed unified framework is applicable to all four cases shown in Figure \ref{fig:structure}. 
Here, we focus on two more complex cases in which $g_\phi$ is unknown, namely, \textit{Case 1: Hysteresis Discovery} and \textit{Case 2: Full Equation Discovery}. 
The remaining two cases can be addressed using analogous procedures within the same framework. 
\revised{For example, under the \textit{Parameter Estimation} setting, where the structures of both $f_\theta$ and $g_\phi$ are known, our framework is only required to estimate the parameters, making it evidently a simpler case.}

\revised{The differences and similarities between these two cases can be seen in Table \ref{tab:cases}. For both cases, $g_\phi$ is assumed to be completely unknown. Therefore, we propose to reformulate the state-space system into a form that does not explicitly involve $dz/dt$:
\begin{equation}
\label{eq:ss_form}
\frac{d}{dt}
\begin{bmatrix}
x \\  
\dot{x}
\end{bmatrix}
=
\begin{bmatrix}
\dot{x} \\  
f_{\theta}(x, \dot{x}, u, z)
\end{bmatrix}
,
\end{equation}
while $z$ itself is treated as trainable thus still retained throughout the model.}

\revised{It should be noted that, since the hysteretic variable $z(t)$ is unobserved and $f_\theta$ is represented by a flexible neural network, part of the hysteretic contribution may in principle be absorbed into the learned motion equation when the available excitation and sensing information are insufficient. In the proposed framework, this ambiguity is practically mitigated by explicitly retaining $z(t)$ as a trainable hysteretic trajectory, embedding the learning process into a solver-based trajectory-matching formulation, and validating the recovered symbolic equations by solving them on testing trajectories. Nevertheless, strict structural identifiability in the fully unconstrained setting is not guaranteed and should be assessed together with prediction accuracy, recovered hysteresis loops, and the physical plausibility of the discovered equations.}

\revised{The key distinction between these two cases lies in whether the structure of $f_\theta$ is known.  
For \textit{Case~1}, assuming that $f_\theta$ is known as in Eq.~\eqref{exp: eq_motion}, but the parameters in it are unknown.
For \textit{Case~2}, since the structure of $f_\theta$ is also unknown, it is represented by a neural network \revised{$\mathcal{N}_\theta(\cdot)$}, parameterized by trainable parameters $\theta$.
More details on network architecture and training can be seen in \ref{Network architecture}.
}

\subsection{Solver-based hysteretic variable learning} \label{step2}
Step 1 of the proposed framework focuses on solver-based hysteretic variable learning, in which the responses to the \revised{reformulated state-space} are simulated using an embedded numerical solver rather than via pointwise regression. Unlike pointwise regression methods that approximate the system's temporal evolution at individual time steps, the solver-based approach enforces temporal consistency by integrating the governing equations over time, thereby capturing the implicit dependence of the hysteretic variable on the system states. \revised{This step is divided into following several parts.}

\begin{revisedblock}
\textbf{Trainable vector variable $\mathbf{z}$.}
In both cases, the structure of the hysteretic link equation ($g_\phi$) is unknown, and the hysteretic variable $\mathbf{z}$ is unobserved. Therefore, we treat $\mathbf{z}$ as a trainable hysteretic vector variable, which is optimized jointly with the model parameters $\theta$.
This design is because $g_\phi$ has a more complex expression, such as non-smooth operators, which is difficult to represent accurately with a neural network, whereas only the value of the vector $\mathbf{z}$ is required in this step. Therefore, treating $\mathbf{z}$ as a trainable parameter provides a more flexible formulation.
\end{revisedblock} 

\begin{revisedblock}
\textbf{Initialization.}
To initialize the model parameters, small-amplitude responses are used to estimate the \revised{initial values} of $\theta$ and $\mathbf{z}$. Table \ref{tab:cases} shows the initialization of two cases.

For \textit{Case 1}, $z(t)$ in Eq.~\eqref{exp: eq_motion} is omitted to simplify dynamic motion equation to:
\begin{equation}
m\ddot{x}(t) + c\dot{x}(t) + kx(t) = u(t).
\end{equation}
Then, the initial values of $m, c, k$ are estimated by the least-squares method, while the initial value $\alpha$ is randomly initialized.

For \textit{Case 2}, the parameters $m, c, k, \alpha$ are obtained in the same way as in \textit{Case 1} for estimating the initial value $\mathbf{z}$, and the neural network parameters $\theta$ are randomly initialized.  

For the initialization of $\mathbf{z} =[z(t_1),z(t_2),\ldots,z(t_N)]^T$, where $N$ denotes the total number of sampled time points, based on Eq.~\eqref{exp: eq_motion}, we can express as:
\begin{equation}
z(t) = [u(t) - m\ddot{x}(t) - c\dot{x}(t) - kx(t)]/\alpha
\end{equation}
and get the initial $\mathbf{z}$.
This procedure provides a proper physically consistent initialization for the learnable hysteretic model, improving convergence and stability of solver-based optimization.
\revised{It should be emphasized that this initialization is used only as an optimization aid, and does not prescribe the final symbolic form of either $f_\theta$ or $g_\phi$. In \textit{Case 2}, $f_\theta$ is still represented by a neural network during solver-based learning, and the final explicit governing equations are obtained only after SR. Therefore, the initialization provides a mild physically motivated prior for stabilizing the ill-posed optimization problem, rather than a hard-coded structural constraint.}
\end{revisedblock}

\begin{revisedblock}
\textbf{Embedded Solver with trainable $\theta$ and $\mathbf{z}$.}
In this step, the governing equations are reformulated into a state-space representation, within which learnable hysteretic model is embedded inside a differentiable ODE solver such as RK4. 

Given the known time stamps $t_i$ ($i = 1,2,\ldots,N$), the time step $\Delta t$, the corresponding input $u_i$, and the initial state $[x(t_1),\dot{x}(t_1)]^T$, the state trajectory can be computed sequentially using the ODE solver as:
\begin{equation}
\begin{bmatrix}
x(t_{i+1}) \\
\dot{x}(t_{i+1})
\end{bmatrix}
=
\mathrm{SolverStep}
\left(
\begin{bmatrix}
x(t_i) \\
\dot{x}(t_i)
\end{bmatrix}
,
\Delta t,\,
\begin{bmatrix}
\dot{x}(t_i) \\
f_{\theta}(x(t_i),\,\dot{x}(t_i),\,u(t_i),\,z(t_i))
\end{bmatrix}
\right)
.
\end{equation}

From above solver step, it can be seen that each element in the vector $\mathbf{z} =[z(t_1),z(t_2),\ldots,z(t_N)]^T$, together with $\theta$, is used within the solver to generate the predicted responses $\mathbf{x}_{\text{pred}}$, $\dot{\mathbf{x}}_{\text{pred}}$, and $\ddot{\mathbf{x}}_{\text{pred}}$, and is continuously updated during the training iterations. In this way, the solver is embedded into the subsequent training process.
\end{revisedblock}

\begin{revisedblock}
\textbf{Training.}
Based on the above embedded solver formulation, the training inputs are given by $\{[x(t_1),\dot{x}(t_1)]^T,\, f_{\theta},\, \mathbf{u},\, \mathbf{z}\}$ ($[x(t_1),\dot{x}(t_1)]^T$ is the initial condition), and the solver produces the predicted responses $\mathbf{x}_{\text{pred}}$, $\dot{\mathbf{x}}_{\text{pred}}$, and $\ddot{\mathbf{x}}_{\text{pred}}$. In both cases, training is performed by backpropagation to update the parameters $\theta$ and $\mathbf{z}$. 

We design a loss function that is flexible to the types of available data. 
Given the known time stamps $t_i, i=1,2,\ldots,N$, the corresponding system responses such as $x(t_i)$, $\dot{x}(t_i)$, and $\ddot{x}(t_i)$ (which may include a subset of them depending on data availability).
As shown in Figure \ref{fig:structure} and \ref{fig:framework}, the formulation is flexible with respect to the available system responses.
The corresponding loss function is:
\begin{equation}
\begin{array}
{@{}l@{}}
\mathcal{L}\text{oss}_{\theta,\mathbf{z}} \bigl( \mathbf{x}_{\text{pred}}, \dot{\mathbf{x}}_{\text{pred}}, \ddot{\mathbf{x}}_{\text{pred}} \bigr)
= \left\lVert\,
\mathbf{S}
\,
\begin{bmatrix}
\mathbf{x}_{\mathrm{pred}}^T \\
\dot{\mathbf{x}}_{\mathrm{pred}}^T \\
\ddot{\mathbf{x}}_{\mathrm{pred}}^T
\end{bmatrix}
- \mathbf{Y}
\right\rVert_{2}^{2}
\end{array}
,
\end{equation}
where the selection matrix $\mathbf{S}$ specifies which quantities are observed and $\mathbf{Y}$ denotes the corresponding available measurements (the same definitions as in the caption of Figure \ref{fig:structure}).

In practice, when some response components are not directly measured, they can be reconstructed from the available signals through smoothing, numerical differentiation, or numerical integration. In our examples, this means that the full set of responses is numerically completed, so that $\mathbf{S}$ is an identity matrix. 
More generally, in our framework, $\mathbf{S}$ can be specified flexibly according to the available observations, because $x(t_i)$, $\dot{x}(t_i)$, and $\ddot{x}(t_i)$ can all be incorporated into the optimization of $\theta$ and $\mathbf{z}$. This is one aspect of the flexibility of the proposed framework.
Once training is completed, the predicted system responses $\mathbf{x}_{\text{pred}}$, $\dot{\mathbf{x}}_{\text{pred}}$, and $\ddot{\mathbf{x}}_{\text{pred}}$, together with the predicted hysteretic variable $\mathbf{z}_{\text{pred}}$, are obtained. 
\end{revisedblock}

Through this step, the hysteretic variable is learned directly from measurements, addressing one of the central difficulties in hysteresis modeling. Importantly, this process avoids restrictive assumptions, such as prescribing a specific functional form for the hysteretic link equation ($g_\phi$) or relying on strongly constrained structural priors, and provides essential data-driven guidance for the subsequent Step 2: Equation discovery, where SR is employed to recover explicit governing equations.

\subsection{Equation discovery}  \label{step3}
\revised{When Step 1 is completed, we can obtain the predicted hysteretic variable $\mathbf{z}_{\text{pred}}$.
The corresponding $\dot{\mathbf{z}}_{\text{pred}}$ can then compute via finite differences, denoted $\mathrm{Diff}(\mathbf{z}_{\text{pred}})$.
It should be noted that the finite difference operation is applied to learned hysteretic trajectories after solver-based training, rather than directly to noisy raw measurements. In our implementation, a smoothing penalty is imposed on the learned $\mathbf{z}_{\text{pred}}$ during training to suppress high-frequency oscillations. Therefore, the derivative computation is performed on smooth-regularized hysteretic trajectories.
For the convenience of narration, all the trajectories we will mention in the following will be written as $\dot{x}_{\text{pred}}$, $z_{\text{pred}}$, and $\dot{z}_{\text{pred}}$, etc.}

Step 2 of the framework is dedicated to extracting interpretable governing equations. For the hysteretic link equation ($g_\phi$), as illustrated in Figure \ref{fig:framework}, symbolic regression (SR) is applied to the learned trajectories of $\dot{x}_{\text{pred}}$, $z_{\text{pred}}$, and $\dot{z}_{\text{pred}}$ to search for mathematical expressions $\dot{z} = g_\phi(\dot{x}, z)$, and this can be written as $\mathrm{SR}\left(\dot{z}_{\text{pred}},\dot{x}_{\text{pred}},z_{\text{pred}}\right)$.
For dynamic motion equation ($f_\theta$), in \textit{Case 1: Hysteresis Discovery}, with known equation structure and the estimated values of $m,c,k,\alpha$, the discovered dynamic motion equation is expressed as $f_\theta = (u - c\dot{x} - kx - \alpha z)/m$; in \textit{Case 2: Full Equation Discovery}, using the same SR method like $g_\phi$, we can obtain the discovered dynamic motion equation $\ddot{x} = f_\theta(x,\dot{x},z,u)$, and this can be written as $\mathrm{SR}\left(\ddot{x}_{\text{pred}},x_{\text{pred}},\dot{x}_{\text{pred}},z_{\text{pred}},u\right)$.

A key advantage of SR lies in its flexibility. Unlike approaches such as SINDy that rely on a predefined functional library, SR enables customization of the search space by incorporating possible operators (e.g.,~absolute value and exponential) and explicit control over expression complexity. In particular, SR allows the specification of operator forms without fixing their exact exponents or coefficients --- for example, one can include $|\cdot|^n$ with $n$ treated as a free constant to be learned. This stands in contrast to SINDy, which requires enumerating many such candidate terms in the predefined library, making it impractical or unfriendly when $n$ is non-integer. Our subsequent experiments will further illustrate this difference. This adaptability mitigates overfitting, promotes interpretability, and facilitates the recovery of compact, physically plausible differential equations that reliably describe the system dynamics.  
\revised{We also acknowledge that SR introduces a mild structural prior through its user-specified operator set and search constraints. In this work, we use a compact operator set $\{+,-,\times,|\cdot|,|\cdot|^{n}\}$ (with $n$ identified as a free constant) together with an explicit complexity penalty and simple grammar constraints (e.g.,~excluding division and disallowing nested absolute/absolute-power compositions) to promote parsimony and reproducibility, which defines the admissible space of discoverable equations.
Additional details on the computational complexity and scalability of the proposed framework are given in \revised{\ref{sec:complexity}}.
}

\section{Experiments}  \label{sec4}
To evaluate the proposed framework, we apply the framework to a series of experiments on representative hysteretic systems. 
These experiments aim to assess both predictive accuracy and generalization ability under varying excitation types, noise conditions and initial conditions.
For the noise conditions part, robustness with respect to measurement uncertainty is examined by adding zero-mean Gaussian white noise to the measured data, with signal-to-noise ratio (SNR) defined as:
\begin{equation}
\label{eq:noise}
\text{SNR}(\text{dB}) = 10 \log_{10} \left( \frac{P_{\text{signal}}}{P_{\text{noise}}} \right),
\end{equation}
where $P_{\text{signal}}$ and $P_{\text{noise}}$ denote the average power of the clean signal and the added noise, respectively. To better demonstrate the performance of the proposed framework, the prediction accuracy is quantified using the normalized root mean square error (NRMSE), defined as:
\begin{equation}
\label{eq:nrmse}
\mathrm{NRMSE}(y_{\mathrm{pred}}, y_{\mathrm{true}}) =
\frac{\sqrt{\frac{1}{N} \sum_{i=1}^{N} (y_{\mathrm{pred},i} - y_{\mathrm{true},i})^2}}
{\max\{y_{\text{true}}\} - \min\{y_{\text{true}}\}},
\end{equation}
where $y_{\mathrm{true},i}$ and $y_{\mathrm{pred},i}$ denote the $i$th observed (ground truth) and predicted values of the target quantity, respectively; $N$ is the total number of samples. 
The normalization by the range of $y_{\mathrm{true}}$ ensures scale-invariant comparison across different datasets or measurement magnitudes.

The subsequent subsections present \revised{four} case studies of increasing complexity: (i) the Bouc-Wen hysteretic system benchmark, (ii) a complex structure with nonlinear and fractional terms, (iii) an experimental single degree-of-freedom (SDOF) yielding structure, and \revised{(iv) Extension to multiple degree-of-freedom (MDOF) systems}. Collectively, these experiments provide a comprehensive evaluation of the proposed framework.

\subsection{Bouc-Wen hysteretic system benchmark}   \label{bouc original}

The Bouc-Wen hysteretic benchmark, developed by No\"el and Schoukens, provides a well-established testbed for evaluating nonlinear system identification methods under hysteretic dynamics~\cite{noel2016hysteretic}. The
benchmark builds on the Bouc-Wen model of hysteresis~\cite{bouc1967forced,wen1976method}, which has been extensively used in structural dynamics and control applications~\cite{ismail2009hysteresis,ikhouane2007systems}. The
benchmark data are generated through accurate integration of the Bouc-Wen equations using a Newmark scheme, including estimation data obtained from noisy multisine excitations for model identification, as well as separate
fixed test datasets for validation.  
In particular, two fixed test datasets are provided: (i) a sinesweep excitation with amplitude of 40 $N$ sweeping from 20--50~Hz and (ii) a random-phase multisine excitation covering the 5--150~Hz frequency
band~\cite{noel2016hysteretic}. These datasets are widely adopted in the system identification community for assessing generalization capabilities of nonlinear models.   

In our experiments, the sinesweep dataset is used for training, while the multisine dataset is reserved for testing. 
This configuration evaluates the ability of the proposed framework to generalize across different excitation types. The goal is not only to capture the nonlinear hysteretic dynamics present in the training data but also to validate robustness and predictive accuracy under unseen forcing conditions.

The Bouc-Wen hysteretic system can be described by a SDOF nonlinear mass--spring-damper oscillator, where the restoring force includes both elastic and hysteretic components. Following the benchmark description
in~\cite{noel2016hysteretic}, the dynamics are governed by the coupled differential equations:
\begin{equation}
\label{bouc-wen benchmark eq}
\left\{
\begin{aligned}
&m\ddot{x}(t) + c\dot{x}(t) + kx(t) + \alpha z(t) = u(t), \\
&\dot{z}(t) = A\dot{x}(t) - \beta |\dot{x}(t)| \,|z(t)| ^{n-1} z(t) - \gamma \dot{x}(t)|z(t)|^{n},
\end{aligned}
\right.
\end{equation}
where $x(t)$ is the displacement response, $u(t)$ is the external excitation, and $z(t)$ is the hysteretic variable. The parameters $m$, $c$, and $k$ denote the mass, damping, and linear stiffness, respectively, while $\alpha$, $A$, $\beta$, $\gamma$, and $n$ control the amplitude, shape, and smoothness of the hysteresis loop. For the benchmark configuration, the parameters are set to $m=2.0, c=10.0, k=50000.0, \alpha=1.0, A=50000.0, \beta=800.0, \gamma=-1100.0$ and $n=1.0$.

\vspace{\baselineskip}
\vspace{\baselineskip}
\vspace{\baselineskip}
\vspace{\baselineskip}

\begin{figure} [!t] 
    \centering
    \includegraphics[width=0.8\linewidth]{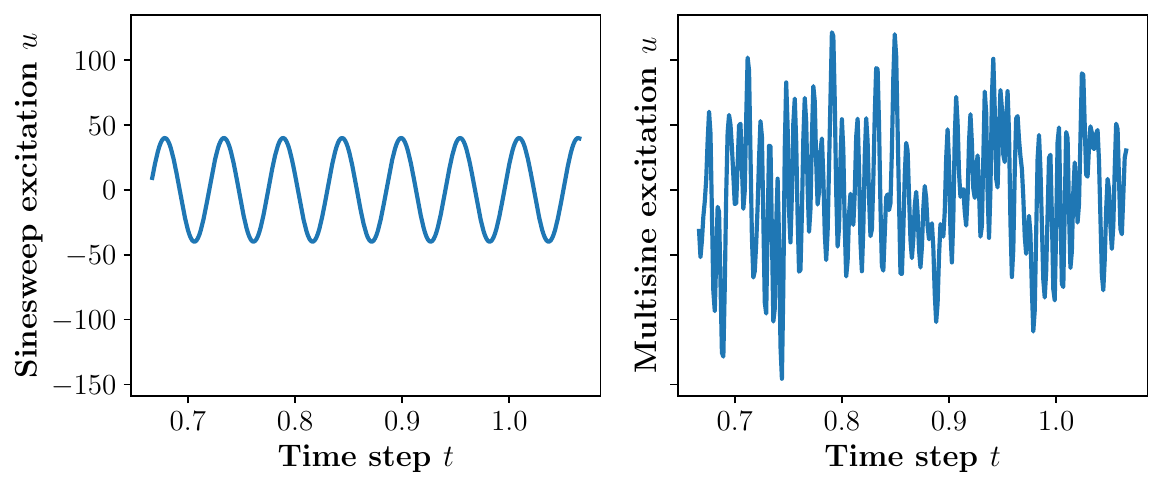}
    \caption{Different types of external excitation of benchmark data. The sinesweep signal (left) is used as the training input (Training and Testing 1), while the multisine signal (right) serves as the testing input (Testing 2).}
    \label{fig:excitation exp1}
\end{figure}

As the first benchmark case, 
\revised{Figure \ref{fig:excitation exp1} illustrates the two external excitations employed for sinesweep signal training and multisine signal testing.}
In addition, we further consider noisy scenarios by adding Gaussian noise with signal-to-noise ratios of 30~dB, 20~dB, and \revised{10} dB, corresponding to low, medium, and high noise levels, respectively.

\begin{figure} [!t]
    \centering
    \includegraphics[width=1.0\linewidth]{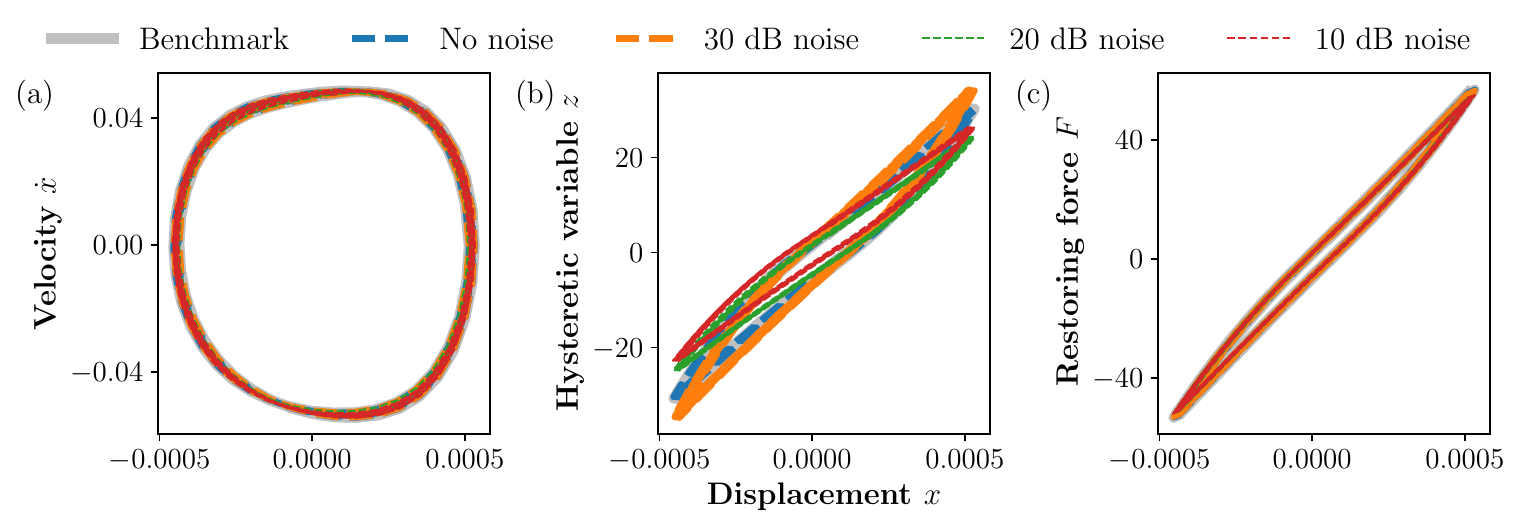}
    \caption{Hysteresis loops of benchmark data. (a) $\dot{x}-x$ hysteresis loop. (b) $z-x$ hysteresis loop. (c) $F-x \; (F = kx + \alpha z)$ hysteresis loop.}
    \label{fig:hysteresis loops exp1}
\end{figure}

\revised{In our framework, the structure of the hysteretic link equation $\dot{z}=g_{\phi}(\dot{x},z)$ is regarded as unknown in both \textit{Case 1: Hysteresis Discovery} and \textit{Case 2: Full Equation Discovery}. In this experiment, this means that the second equation in Eq.~\eqref{bouc-wen benchmark eq} is not assumed a priori. Accordingly, the subsequent results are reported as ``discovered equations'', rather than merely ``parameter estimation''. We do not prescribe a classical or generalized Bouc-Wen functional form for the hysteretic evolution. Instead, the hysteretic variable trajectory is first learned through the solver-based scheme, and symbolic regression (SR) is then used to recover an explicit symbolic expression for the hysteretic link.}
Here, we assume to know the structure of the dynamics motion equation, thus it is a problem of \textit{Hysteresis Discovery}. 
Figure \ref{fig:hysteresis loops exp1} shows the hysteresis loops learned by the proposed framework, in terms of three different representations. It can be observed that, even in the presence of strong noise, the proposed model successfully captures nonlinear memory-dependent hysteretic behaviors.  

We define two different types of testing cases. 
In Testing 1, the external excitation is the same as in Training, but the data are taken from a time span beyond the training window. 
In Testing 2, an \revised{unseen} external excitation is applied, providing an out-of-distribution scenario relative to training. 
  
For a quantitative evaluation, Table \ref{tab:equation exp1} and \ref{tab:nrmse exp1} report the performance in terms of discovered equations and NRMSE at different noise levels, respectively. The results show that the proposed framework consistently achieves an accurate reconstruction of explicit governing equations, with the responses reproduced from the discovered explicit equations exhibiting low NRMSE values. 

\begin{table} [!b]
\centering
\caption{Discovered equations results of benchmark data ($t$ is omitted for brevity).} 
\resizebox{0.75\textwidth}{!}{%
\begin{tabular}{@{}c c@{}}
\revised{\textbf{True equations}} & 
\makecell[l]{
    \revised{$2.0000 \ddot{x} + 10.0000 \dot{x} + 50000.0000 x + 1.0000 z = u$} \\
    \revised{$\dot{z} = 50000.0000 \dot{x} - 800.0000|\dot{x}| z + 1100.0000 \dot{x} |z|$} 
    } \\ [12pt]
\toprule
\textbf{Noise} & 
\makecell[c]{\textbf{Discovered equations} \\
}\\
\midrule
\makecell[c]{Noise-free} 
&  
\makecell[l]{%
    $1.9998 \ddot{x} + 10.0231 \dot{x} + 49893.8617 x + 1.0110 z = u$\\
    $\dot{z} = 49633.1496 \dot{x} - 779.1857 |\dot{x}| z + 1049.3318 \dot{x} |z|$} \\
\midrule
\makecell[c]{30dB} 
& 
\makecell[l]{%
    $1.8623 \ddot{x} + 10.2269 \dot{x} + 47507.9528 x + 1.0210 z = u$\\
    $\dot{z} = 50858.0627 \dot{x} - 734.82275 |\dot{x}| z + 922.7398 \dot{x} |z|$} \\
\midrule
\makecell[c]{20dB} 
&  
\makecell[l]{%
    $1.7976 \ddot{x} + 12.8240 \dot{x} + 46127.6819  x + 1.2557 z = u$\\
    $\dot{z} = 43574.2517 \dot{x} - 727.0708 |\dot{x}| z + 727.0708 \dot{x} |z|$} \\
\midrule
\makecell[c]{\revised{10dB}} 
&  
\makecell[l]{%
    \revised{$1.6149 \ddot{x} + 14.8180 \dot{x} + 45758.4897  x + 1.4018 z = u$}   \\
    \revised{$\dot{z} = 402647.4894 \dot{x} - 700.5687 |\dot{x}| z + 700.5687 \dot{x} |z|$}
    }    \\
\bottomrule
\end{tabular}
}
\label{tab:equation exp1}
\end{table}

\begin{table} [!h] 
\centering
\caption{NRMSE results of benchmark data.} 
\resizebox{0.7\textwidth}{!}{
\begin{tabular}{@{}c c c c c@{}}
\toprule
\textbf{Noise} & \textbf{Responses} & 
\makecell[c]{\textbf{Training} \\ \textbf{NRMSE}}  & 
\makecell[c]{\textbf{Testing 1} \\ \textbf{NRMSE}} & 
\makecell[c]{\textbf{Testing 2} \\ \textbf{NRMSE}} \\
\midrule
\multirow{2}{*}{Noise-free} 
& Displacement $x$   & 0.15\%  & 0.13\%  & 0.93\% \\
& Velocity $\dot{x}$ & 0.44\%  & 0.49\%  & 1.06\% \\
\midrule
\multirow{2}{*}{30 dB} 
& Displacement $x$   & 0.85\%  & 0.78\%  & 0.67\% \\
& Velocity $\dot{x}$ & 1.29\%  & 1.13\%  & 0.96\% \\
\midrule
\multirow{2}{*}{20 dB} 
& Displacement $x$   & 2.41\%  & 1.64\%  & 3.30\% \\
& Velocity $\dot{x}$ & 3.08\%  & 2.15\%  & 3.37\% \\
\midrule
\multirow{2}{*}{\revised{10 dB}} 
& \revised{Displacement $x$}   & \revised{4.80\%}  & \revised{4.52\%}  & \revised{5.11\%} \\
& \revised{Velocity $\dot{x}$} & \revised{5.57\%}  & \revised{5.90\%}  & \revised{6.13\%} \\
\bottomrule
\end{tabular}
}
\label{tab:nrmse exp1}
\end{table}

\begin{figure} [!t]
    \centering
    \includegraphics[width=1.0\linewidth]{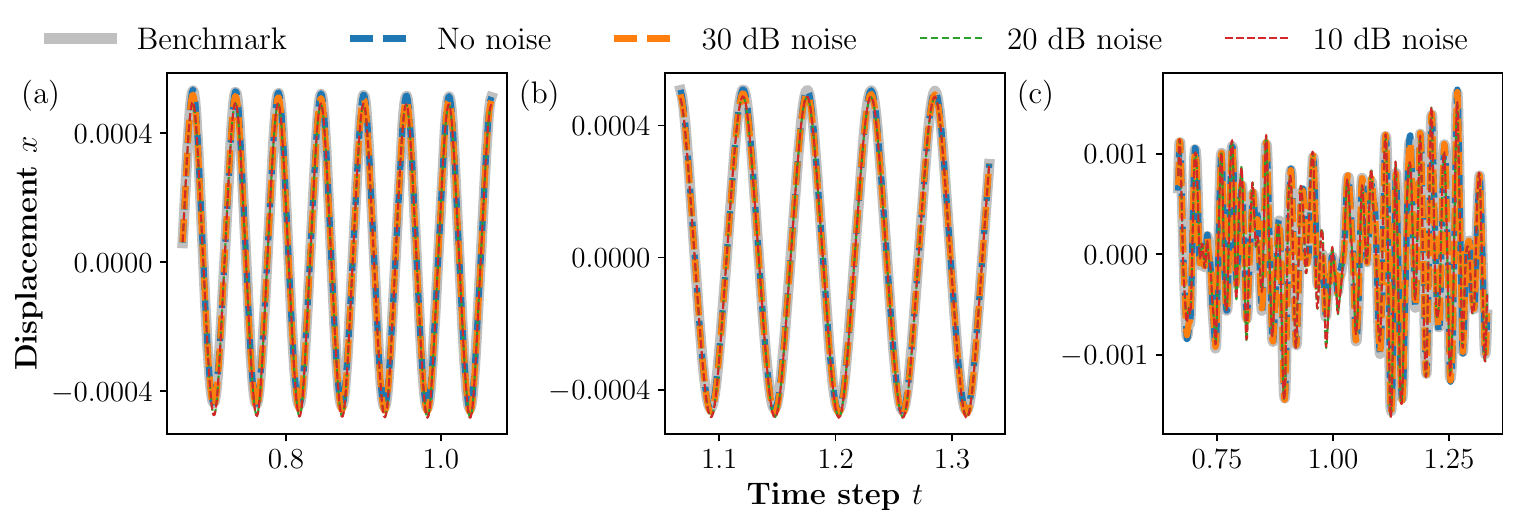}
    \caption{Displacement results of benchmark data. (a) Training results. (b) Testing 1 (same excitation) results. (c) Testing 2 (\revised{unseen} excitation) results.}
    \label{fig:displacement results exp1}
\end{figure}

\begin{figure} [!h]
    \centering
    \includegraphics[width=1.0\linewidth]{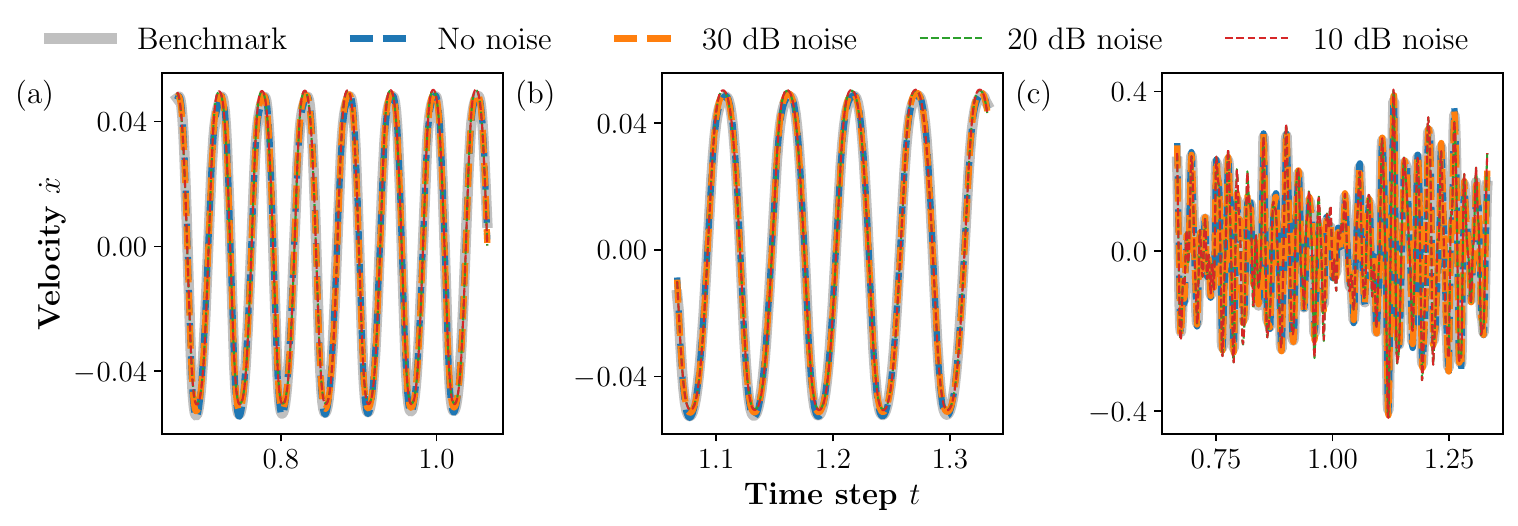}
    \caption{Velocity results of benchmark data. (a) Training results. (b) Testing 1 (same excitation) results. (c) Testing 2 (\revised{unseen} excitation) results.}
    \label{fig:velocity results exp1}
\end{figure}

To further evaluate the validity of the discovered models, \revised{the displacement and velocity responses obtained by numerically solving the discovered governing equations are compared with the corresponding ground truth results in  Figure \ref{fig:displacement results exp1} and \ref{fig:velocity results exp1}, respectively.}
This procedure more directly highlights the effectiveness of equation discovery.
This benchmark experiment demonstrates that the unified framework is capable of reliably identifying hysteretic dynamics, even under noisy conditions. These findings validate the effectiveness of the proposed methodology for equation discovery and dynamic prediction of nonlinear hysteretic systems.

\FloatBarrier

\begin{revisedblock}
To further validate the proposed framework, we also conduct some additional benchmark experiments beyond the baseline setting:
\textit{Cross-excitation validation}, 
\textit{Extrapolation test on unseen amplified excitation}, 
\textit{Noisy wideband random excitation training} 
and \textit{Without hysteresis}.
The detailed context can be seen in \ref{bouc add}.
\end{revisedblock}

\subsection{A complex structure with high-order nonlinear and fractional terms}   

The second experiment considers a more complex hysteretic system to further evaluate the flexibility of the proposed framework. In this case, the restoring force incorporates a cubic nonlinearity in the displacement response, and the hysteretic evolution equation involves a fractional exponent $n$, thereby introducing additional challenges compared to the standard Bouc-Wen formulation. The governing equations are given by:
\begin{equation}
\label{complex_structure} 
\left\{
\begin{aligned}
&m \ddot{x}(t) + c \dot{x}(t) + kx^3(t) + \alpha z(t) = u(t), \\
&\dot{z}(t) = A \dot{x}(t) - \beta \lvert \dot{x}(t) \rvert \lvert z(t) \rvert^{\,n-1} z(t) - \gamma \dot{x}(t) \, \lvert z(t) \rvert^{\,n},
\end{aligned}
\right.
\end{equation}
with system parameters $m=1.0, c=0.8, k=0.5, \alpha=1.0, A=4.0, \beta=5.0$, and $\gamma=-4.0$. Unlike the benchmark case, here we let $n = 1.5$ (fractional), which modifies the smoothness and sharpness of the hysteretic transition and better reflects diverse hysteretic behaviors. 

This experiment highlights the unified nature of the framework, as it remains applicable even when the underlying system exhibits higher-order nonlinearities and non-integer hysteretic exponents.
\revised{In this experiment, we consider the two cases: \textit{Hysteresis Discovery} and \textit{Full Equation Discovery}.}

\begin{figure}[!h]
    \centering
    \captionsetup{labelfont={color=black}, textfont={color=black}}
    \includegraphics[width=1.0\linewidth]{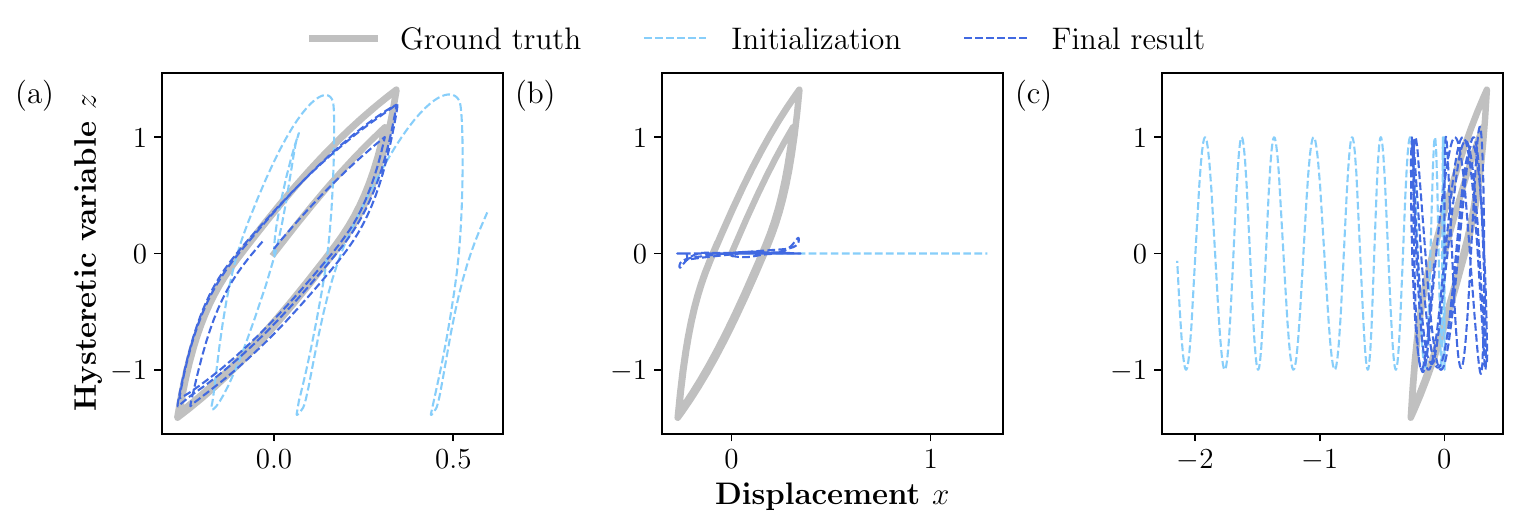}
    \caption{$z$--$x$ hysteresis loops under the same single initial condition and excitation in the \textit{Full Equation Discovery} setting of complex structure data. Three different strategies: (a) Ours: initialization described in \ref{step2}. (b) Zero: $z(t) = 0$. (c) Sine: $z(t) = \sin(2\pi t)$.}
    \label{fig:init exp2}
\end{figure}

\begin{figure}[!h]
    \centering
    \captionsetup{labelfont={color=black}, textfont={color=black}}
    \includegraphics[width=1.0\linewidth]{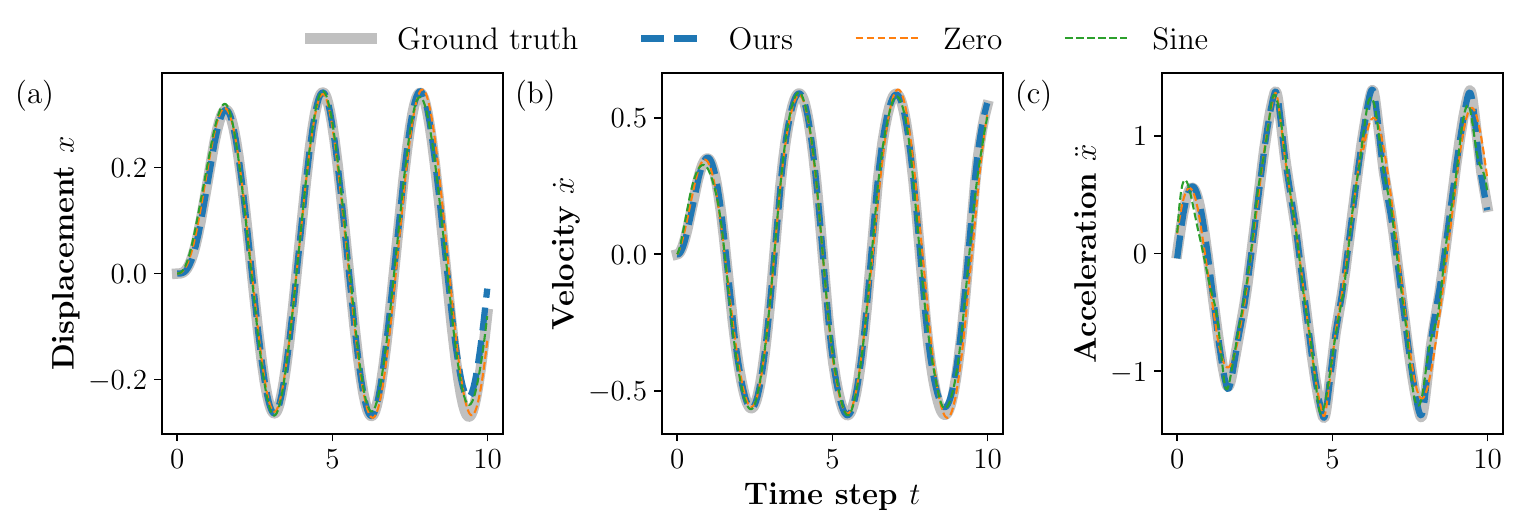}
    \caption{Predicted response results in Step 1 of complex structure data. (a) Displacement. (b) Velocity. (c) Acceleration.}
    \label{fig:init pred exp2}
\end{figure}

\begin{revisedblock}
Before presenting the main experimental results, we first examine the sensitivity of the learned hysteretic variable to initialization in the most challenging \textit{Full Equation Discovery} setting.
Specifically, we consider a fixed training case with only one initial condition and one excitation, while initializing the trainable hysteretic variable $\mathbf{z}$ through Step 1 in three different strategies: 
(a) Ours: initialization described in \ref{step2},
(b) Zero: $z(t) = 0$,
(c) Sine: $z(t) = \sin(2\pi t)$. 
Figure \ref{fig:init exp2} shows the $x$--$z$ hysteresis loops for these three strategies at initialization and after convergence.
Figure \ref{fig:init pred exp2} shows the Step 1 predicted results, it can be seen that all three initialization strategies can accurately predict $x$, $\dot{x}$, and $\ddot{x}$. 
\revised{In this regard, to further verify whether the discovered equations are legitimate, one can implement the discovered equation for predicting the response from a test dataset.}

However, the strategies Zero and Sine struggle to recover the correct hysteresis loops, indicating that accurate response prediction does not necessarily imply accurate learning of the hysteretic variable.
Although the initial hysteresis loop delivered by our initialization strategy differs from the ground truth, it captures hysteretic features compared with the other two strategies. After training, the final optimized results become much closer to the ground truth.

This result suggests that, when the training data are not sufficiently informative, strict uniqueness of the learned hysteretic variable cannot be asserted. 
\revised{Therefore, accurate response prediction is a necessary but not sufficient condition for recovering the hysteretic variable.}
Nevertheless, the proposed solver-based learning scheme remains a practically effective initialization strategy. 
This example represents a relatively extreme case, as only a single training trajectory is used. It also highlights the importance of sufficient diversity in the training data, which is subsequently examined in this experiment by considering multiple initial conditions.
\end{revisedblock}

\revised{In this experiment, the training data are generated from five different initial conditions, while an additional unseen initial condition is used for testing to evaluate the generalizability of the proposed framework.}
The external excitation is defined as $u(t) = \sin(2t)$, representing a smooth periodic input suitable for a dynamic system. The sampling frequency is set to $f_s = 100~\text{Hz}$, and the total simulation duration is $T = 6$ s. Such a setup allows for a comprehensive assessment of the model's robustness under varying initial states while maintaining consistent excitation characteristics.

\begin{revisedblock}
To compare the proposed framework with a representative sparse regression pipeline, we use SINDy~\cite{brunton2016discovering}, which discovers governing equations by performing sparse regression over a pre-defined
candidate function library. 
In addition to the vanilla SINDy, we include two advanced SINDy methods to provide a fairer comparison under strong nonlinearity and noisy measurements. Specifically, we consider parallel implicit SINDy
(SINDy-PI)~\cite{kaheman2020sindy}, which identifies implicitly defined governing relations and leverages multiple trajectories to improve identifiability, and Weak SINDy~\cite{messenger2021weak}, which performs
identification in an integral (weak) form to reduce sensitivity to derivative noise. 
For all SINDy-based methods, we use a candidate library design that is made as consistent as possible with our method, and we tune the key sparsity-related hyperparameters on the training set to avoid biasing the comparison. Specifically, the dynamic motion equation is assumed to be a linear combination of $x$, $\dot{x}$, $z$, and $u$, whereas the hysteretic link equation is constrained to include absolute-value terms and their integer powers from 1 to 5, thereby providing SINDy with sufficient expressive capacity to capture nonlinear hysteretic behaviors.
\end{revisedblock}

\begin{figure}[!t] 
    \centering
    \includegraphics[width=1.0\linewidth]{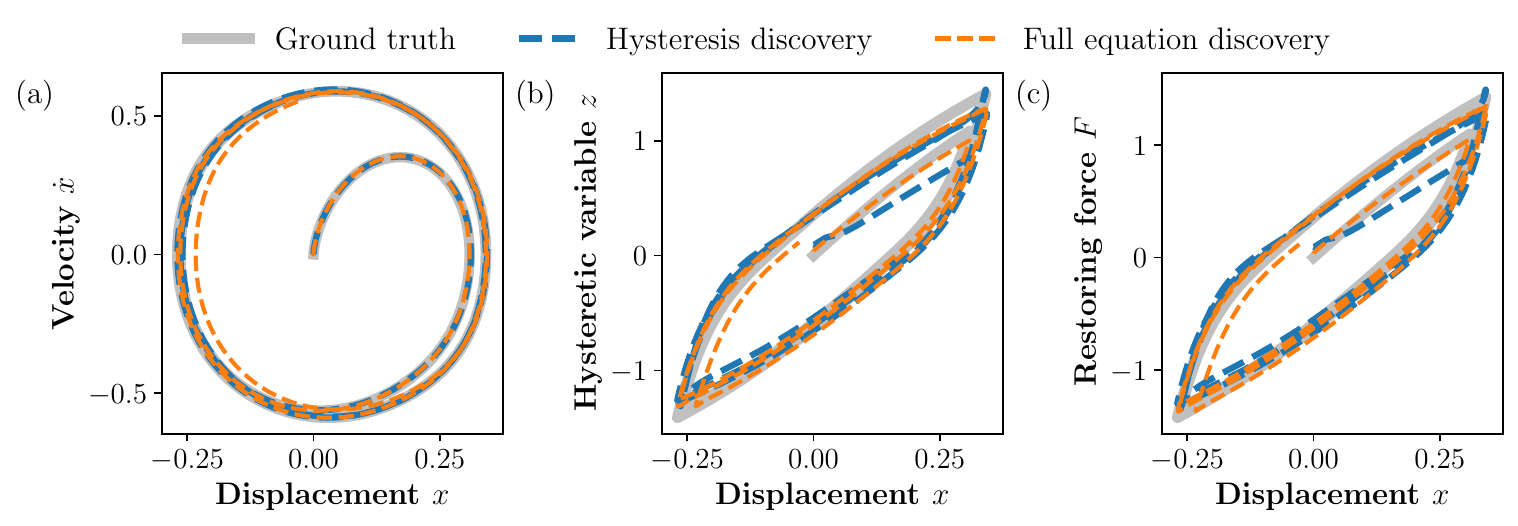}
    \caption{Hysteresis loops of complex structure data. (a) $\dot{x}-x$ hysteresis loop. (b) $z-x$ hysteresis loop. (c) $F-x \; (F = kx^3 + \alpha z)$ hysteresis loop.}
    \label{fig:hysteresis loops exp2}
\end{figure}

Figure \ref{fig:hysteresis loops exp2} shows the hysteresis loops learned by the proposed framework. Both configurations can capture qualitative hysteretic behavior, yet the \textit{Hysteresis Discovery} model clearly achieves better accuracy in reconstructing the loop dynamics. Table \ref{tab:equation exp2} lists the explicit forms of the discovered equations, and Table \ref{tab:nrmse exp2} reports the NRMSE results \revised{for all methods.} 
\revised{It can be seen that Weak SINDy improves the response-level accuracy over vanilla SINDy, most notably under \textit{Hysteresis Discovery}, whereas SINDy-PI provides limited gains and becomes less reliable in \textit{Full Equation Discovery}, where the recovered dynamics can deviate substantially from the true responses.} 

\begin{figure} [!h]
    \centering
    \includegraphics[width=1.0\linewidth]{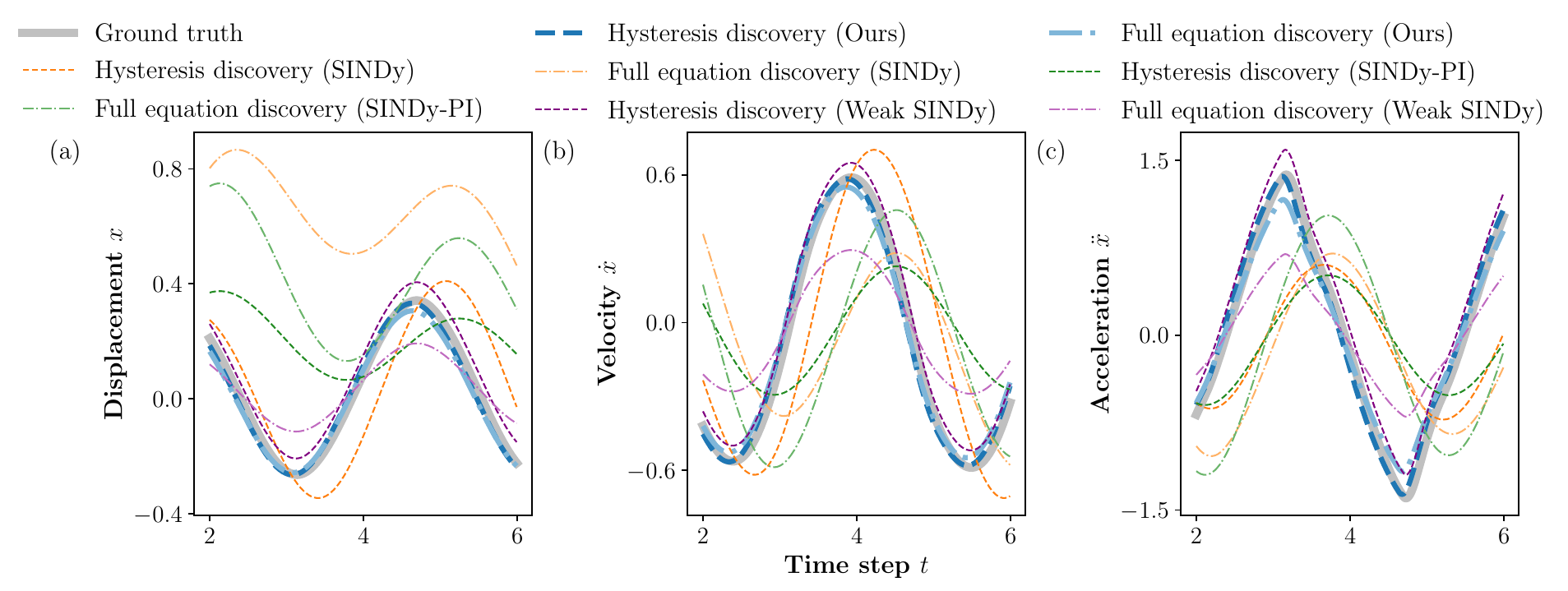}
    \caption{Training results of complex structure data. (a) Displacement. (b) Velocity. (c) Acceleration.}
    \label{fig:training results exp2}
\end{figure}

\begin{figure} [!h]
    \centering
    \includegraphics[width=1.0\linewidth]{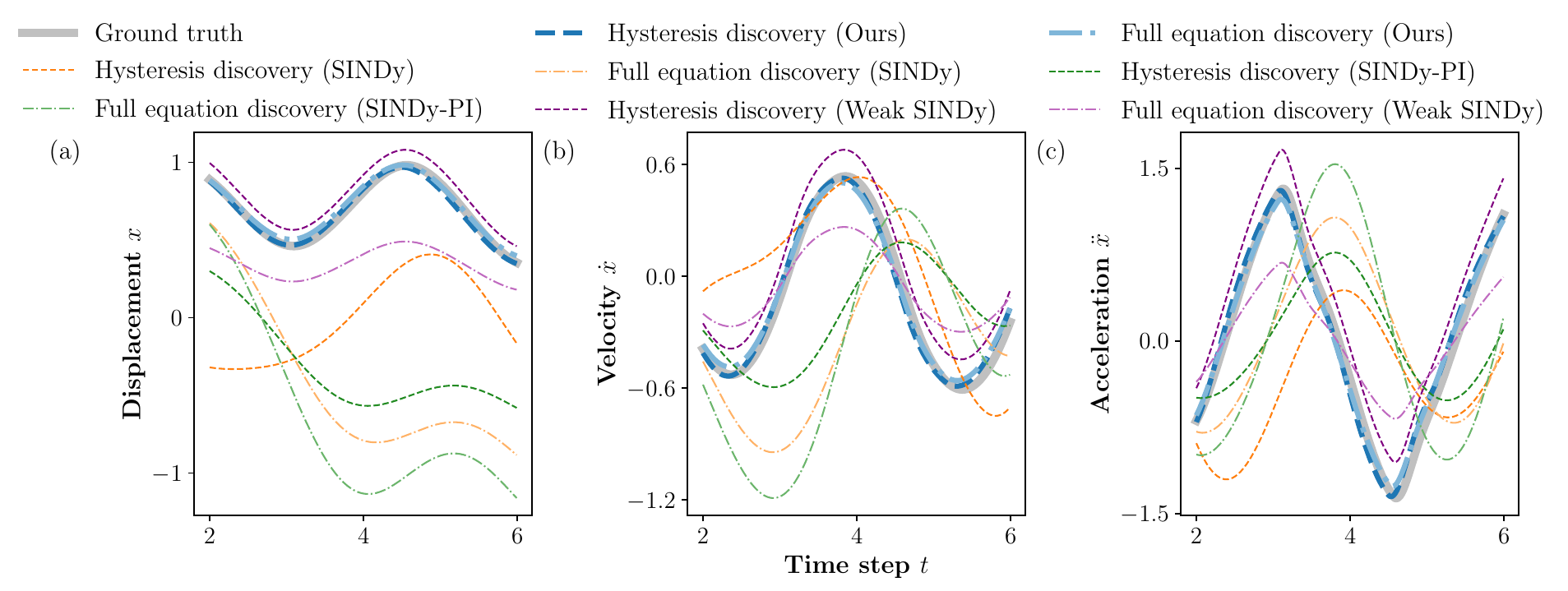}
    \caption{Testing results of complex structure data. (a) Displacement. (b) Velocity. (c) Acceleration.}
    \label{fig:testing results exp2}
\end{figure}

Figure \ref{fig:training results exp2} and \ref{fig:testing results exp2} compare the displacement, velocity, and acceleration responses obtained by solving the discovered equations, compared with the corresponding ground truth responses in the training and testing datasets, respectively. 
\revised{Despite the partial improvement brought by Weak SINDy, both SINDy-PI and Weak SINDy remain constrained by the predefined integer-order candidate library,  and therefore fail to recover the fractional-order hysteretic terms. As shown in Table \ref{tab:equation exp2}, they tend to introduce spurious integer-power terms as compensation. By contrast, our method consistently identifies fractional exponents close to the ground truth and achieves the lowest errors in both training and testing (Table \ref{tab:nrmse exp2}), with \textit{Hysteresis Discovery} remaining more accurate than \textit{Full Equation Discovery} due to the benefit of incorporating prior structural knowledge of dynamics motion equation ($f_\theta$).} 
From these results, it is evident that the proposed framework consistently outperforms SINDy-based methods, achieving lower errors and more faithful recovery of the governing equations with fractional exponents. 

\begin{table}[!t] 
\centering
\caption{Discovered equations results of complex structure data ($t$ is omitted for brevity). SINDy fails to recover the fractional-order terms and instead introduces spurious integer-order terms to compensate.}
\resizebox{1.0\textwidth}{!}{%
\begin{tabular}{@{}c c@{}}
\revised{\textbf{True equations}} & 
\makecell[l]{
    \revised{$1.0000 \ddot{x} + 0.8000 \dot{x} + 0.5000 x^3 + 1.0000 z = 1.0000 u$} \\
    \revised{$\dot{z} = 4.0000 \dot{x} - 5.0000 |\dot{x}| |\revised{z}|^{0.5} z + 4.0000 \dot{x} |z|^{1.5}$} 
    } \\ [12pt]
\toprule
\textbf{Type and method} &
\makecell[c]{\textbf{Discovered equations} 
} \\
\midrule
{\makecell[c]{\textbf{\textit{Hysteresis Discovery}} \\ \textbf{(Ours)}}}  &   
{\makecell[l]{%
    \boldmath
    $0.9388 \ddot{x} + 0.7934 \dot{x} + 0.5366 x^3 + 1.0320 z = u$\\
    \boldmath
    $\dot{z} = 4.0633 \dot{x} - 4.8310 |\dot{x}| |\revised{z}|^{0.5486} z + 3.8942 \dot{x} |z|^{1.5486}$
    }} \\
\midrule
{\makecell[c]{\textit{Hysteresis Discovery} \\ (SINDy)}}  &
{\makecell[l]{%
    $\ddot{x} + 0.8445 \dot{x} + 0.5679 x^3 + 1.0992 z = 1.0649 u$\\
    $\dot{z} = -6.80922 \, |\dot{x}| \, |z| z + 5.00018 \, |\dot{x}| \, |z|^{2} z + 4.98652 \, \dot{x} \, |z|^{2} + 3.73699 \, \dot{x}$}} \\
\midrule
{\makecell[c]{\textit{\revised{Hysteresis Discovery}} \\ \revised{(SINDy-PI)}}}  &
{\makecell[l]{%
    \revised{$0.9097 \ddot{x} + 0.7685 \dot{x} + 0.5165 x^3 + 1.0000 z = 0.9689 u$} \\
    \revised{$\dot{z} =
    -|\dot{x}|\,|z|^{2}z
    +0.4263\,|\dot{x}|\,|z|z
    -0.0583\,|\dot{x}|\,z
    +0.0258\,\dot{x}\,|z|^{2}$}}} \\
\midrule
{\makecell[c]{\textit{\revised{Hysteresis Discovery}} \\ \revised{(Weak SINDy)}}}  &
{\makecell[l]{%
    \revised{$\ddot{x} + 0.8483 \dot{x} + 0.5715 x^3 + 1.1007 z = 1.0678 u$} \\
    \revised{$\dot{z}=
    17.7907\,\dot{x}\,|z|^{2}
    -6.4668\,|\dot{x}|\,|z|\,z
    +3.9354\,\dot{x}
    -1.9870\,|\dot{x}|\,z$}}} \\
\midrule
{\makecell[c]{\textbf{\textit{Full Equation}} \\ \textbf{\textit{Discovery}} \textbf{(Ours)}}}  &
{\makecell[l]{%
    \boldmath
    $\ddot{x} + 0.7689 \dot{x} + 0.4573 x^3 + 1.0423 z = 0.9503u$\\
    \boldmath
    $\dot{z} = 4.0543 \dot{x} - 5.1348 |\dot{x}| |\revised{z}|^{0.6225} z + 3.8224 \dot{x} |z|^{1.6225}$}} \\
\midrule
{\makecell[c]{\textit{Full Equation} \\ \textit{Discovery} (SINDy)}}  &
{\makecell[l]{%
    $\ddot{x} + 0.7490 \dot{x} + 0.5833 x^{3} + 1.2145 z = 0.8992 u$\\
    $\dot{z} = -12.704|\dot{x}||z|z + 6.118|\dot{x}||z|^{2}z + 2.941\dot{x}|z|^{2} - 0.873\dot{x}$}} \\
\midrule
{\makecell[c]{\textit{\revised{Full Equation}} \\ \revised{\textit{Discovery} (SINDy-PI)}}}  &
{\makecell[l]{%
    \revised{$0.0330 \ddot{x} + 0.0601 \dot{x} + x^3 + 0.0555 z = 0.0082 u$} \\
    \revised{$\dot{z} =
    -|\dot{x}|\,|z|^{2}z
    +0.4346\,|\dot{x}|\,|z|\,z
    -0.0602\,|\dot{x}|\,z
    -0.0108\,\dot{x}\,|z|^{2}$}}} \\
\midrule
{\makecell[c]{\textit{\revised{Full Equation}} \\ \revised{\textit{Discovery} (Weak SINDy)}}}  &
{\makecell[l]{%
    \revised{$\ddot{x} + 0.0826 \dot{x} + 0.84054 x^3 + 1.2282 z = 0.6778 u$} \\
    \revised{$\dot{z} =
    -2.6871\,|\dot{x}|\,|z|\,z
    +5.5021\,\dot{x}\,|z|
    -3.6203\,\dot{x}\,|z|^{2}
    +3.2106\,\dot{x}$}}} \\
\bottomrule
\end{tabular}
}
\label{tab:equation exp2}
\end{table}

\begin{table}[!h]
\centering
\caption{NRMSE results of complex structure data.}
\resizebox{0.95\textwidth}{!}{%
\begin{tabular}{@{}c c c c c@{}}
\toprule
\textbf{Type} & \textbf{Responses} & \textbf{Method} &
\makecell[c]{\textbf{Training NRMSE}} &
\makecell[c]{\textbf{Testing NRMSE}} \\
\midrule
\multirow{12}{*}{\makecell[c]{\textit{Hysteresis}\\\textit{Discovery}}}
& \multirow{4}{*}{Displacement $x$}
& \textbf{Ours}        & \textbf{4.21\%} & \textbf{3.84\%} \\
&  & SINDy       & 25.42\% & 36.87\% \\
&  & \revised{SINDy-PI}   & \revised{25.48\%} & \revised{28.37\%} \\
&  & \revised{Weak SINDy} & \revised{7.77\%} & \revised{10.19\%} \\
\cmidrule(lr){2-5}
& \multirow{4}{*}{Velocity $\dot{x}$}
& \textbf{Ours}        & \textbf{3.91\%} & \textbf{4.01\%} \\
&  & SINDy       & 24.79\% & 39.50\% \\
&  & \revised{SINDy-PI}   & \revised{20.56\%} & \revised{26.73\%} \\
&  & \revised{Weak SINDy} & \revised{8.03\%} & \revised{12.06\%} \\
\cmidrule(lr){2-5}
& \multirow{4}{*}{Acceleration $\ddot{x}$}
& \textbf{Ours }       & \textbf{3.86\%} & \textbf{3.88\%} \\
&  & SINDy       & 25.70\% & 40.67\% \\
&  & \revised{SINDy-PI}   & \revised{20.07\%} & \revised{27.66\%} \\
&  & \revised{Weak SINDy} & \revised{7.79\%} & \revised{12.43\%} \\
\midrule
\multirow{12}{*}{\makecell[c]{\textit{Full Equation}\\\textit{Discovery}}}
& \multirow{4}{*}{Displacement $x$}
& \textbf{Ours}        & \textbf{5.28\%} & \textbf{4.57\%} \\
&  & SINDy       & 25.73\% & 37.76\% \\
&  & \revised{SINDy-PI}   & \revised{22.92\%} & \revised{41.41\%} \\
&  & \revised{Weak SINDy} & \revised{15.82\%} & \revised{21.68\%} \\
\cmidrule(lr){2-5}
& \multirow{4}{*}{Velocity $\dot{x}$}
& \textbf{Ours}        & \textbf{4.56\%} & \textbf{4.25\%} \\
&  & SINDy       & 26.56\% & 32.43\% \\
&  & \revised{SINDy-PI}   & \revised{21.97\%} & \revised{37.71\%} \\
&  & \revised{Weak SINDy} & \revised{18.86\%} & \revised{20.55\%} \\
\cmidrule(lr){2-5}
& \multirow{4}{*}{Acceleration $\ddot{x}$}
& \textbf{Ours}        & \textbf{4.86\%} & \textbf{4.31\%} \\
&  & SINDy       & 23.00\% & 36.15\% \\
&  & \revised{SINDy-PI}   & \revised{22.02\%} & \revised{32.64\%} \\
&  & \revised{Weak SINDy} & \revised{17.24\%} & \revised{20.80\%} \\
\bottomrule
\end{tabular}
}
\label{tab:nrmse exp2}
\end{table}

\FloatBarrier

\subsection{A SDOF experimental structure}

\begin{figure}[!t] 
    \centering
    \includegraphics[width=0.9\linewidth]{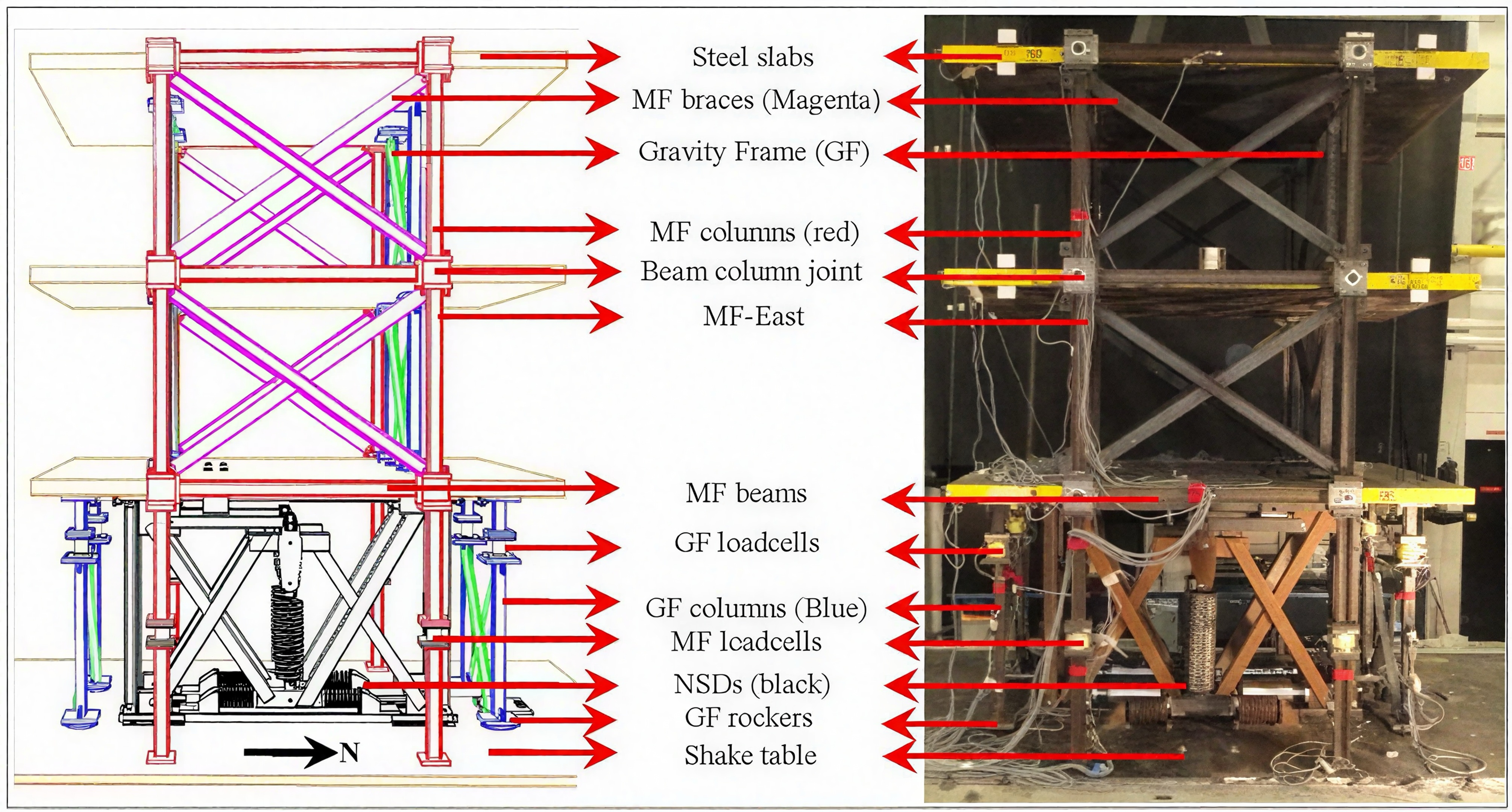}
    \caption{Shaking table test of a braced 3-story structure (equivalent to a SDOF system), with a negative stiffness device (NSD) installed on the first floor~\cite{lai2019sparse}.}
    \label{fig:experiment exp3}
\end{figure}

In this experiment, a shear-type SDOF yielding structure excited by shake-table ground motions is implemented to investigate the proposed framework method. 
The experimental setup of the braced 3-story structure (equivalent to a SDOF system), with and without a negative stiffness device (NSD) \cite{sarlis2013negative,pasala2013adaptive} is illustrated in Figure \ref{fig:experiment
exp3}~\cite{lai2019sparse,pasala2013seismic,Attary17022015}, where the NSD is installed on the first floor to alter nonlinear behaviors.

We adopt the governing equation structure $m \ddot{x}(t) + c \dot{x}(t) + k x(t) + \alpha z(t) = -m u(t) $, where $x(t)$ is the displacement, $\dot{x}(t)$ and $\ddot{x}(t)$ are its velocity and acceleration, $u(t)$ is the measured base acceleration, and $z(t)$ is the hysteretic variable. This setup belongs to the \textit{hysteresis discovery} type, in which the structure of the equation of motion is predefined (which needs to identify unknown system parameters), and the nonlinear hysteretic link equation is assumed to be unknown. 

We reuse the real shake table experiment from~\cite{lai2019sparse}, and study the structure in two scenarios: (i) \textit{without} an NSD and (ii) \textit{with} an NSD installed on the first floor. Following the setup
in~\cite{lai2019sparse}, we use the measured displacement $x(t)$, the computed velocity $\dot{x}(t)$ (via differentiation and smoothing), the measured acceleration $\ddot{x}(t)$, and the shake-table excitation
$u(t)$ as inputs to our framework. For the \textit{without NSD} case, a type-I 60\% scaled Pacoima earthquake is used for training, and a reversed type-II 60\% Pacoima record for testing. For the \textit{with NSD}
case, the training signal concatenates 43\% and 65\% scaled Pacoima segments, while the testing input is a 55\% scaled Sylmar record (see Figs 23--25 of~\cite{lai2019sparse} for details).

\begin{figure} [!t]
    \centering
    \includegraphics[width=0.8\linewidth]{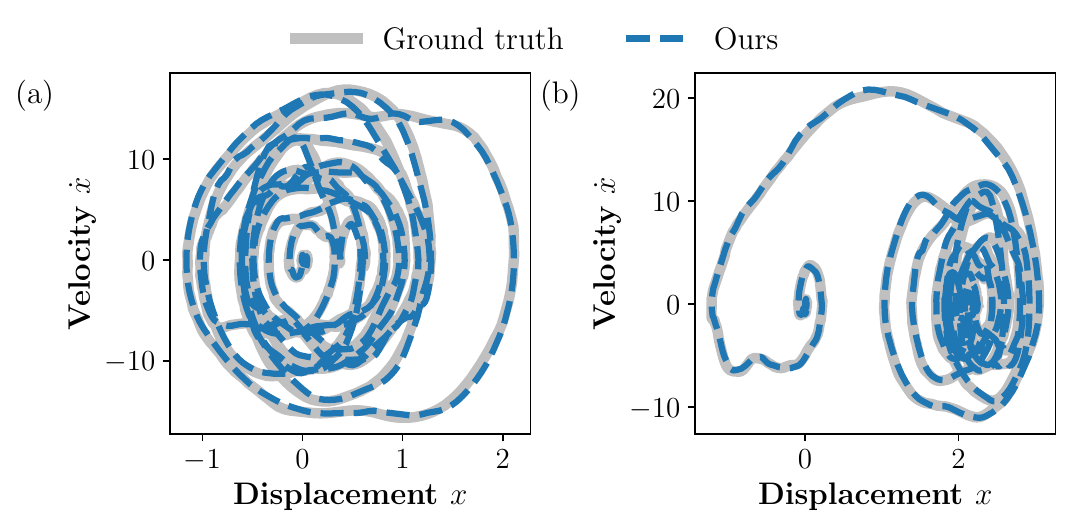}
    \caption{$\dot{x}-x$ hysteresis loops of SDOF experimental structure. (a) Without NSD. (b) With NSD.}
    \label{fig:hysteresis loops exp3}
\end{figure}

\begin{figure}  [!h]
    \centering
    \includegraphics[width=1.0\linewidth]{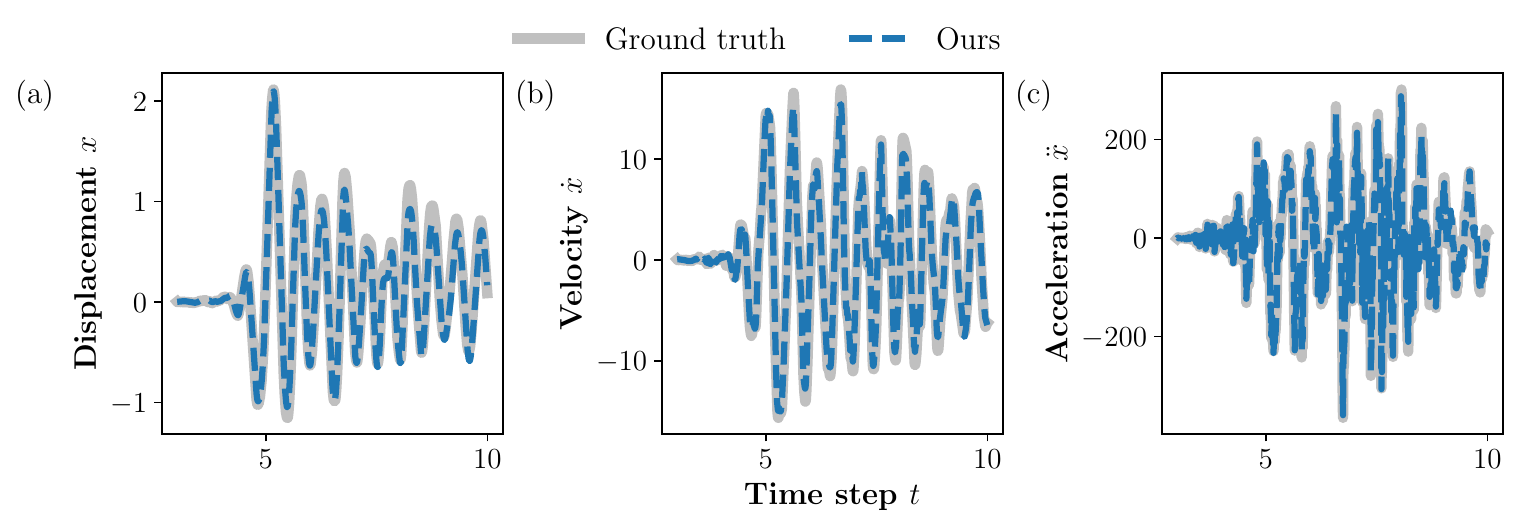}
    \caption{Training results of SDOF experimental structure data without NSD. (a) Displacement. (b) Velocity. (c) Acceleration.}
    \label{fig:training results without exp3}
\end{figure}

\begin{figure} [!h]
    \centering
    \includegraphics[width=1.0\linewidth]{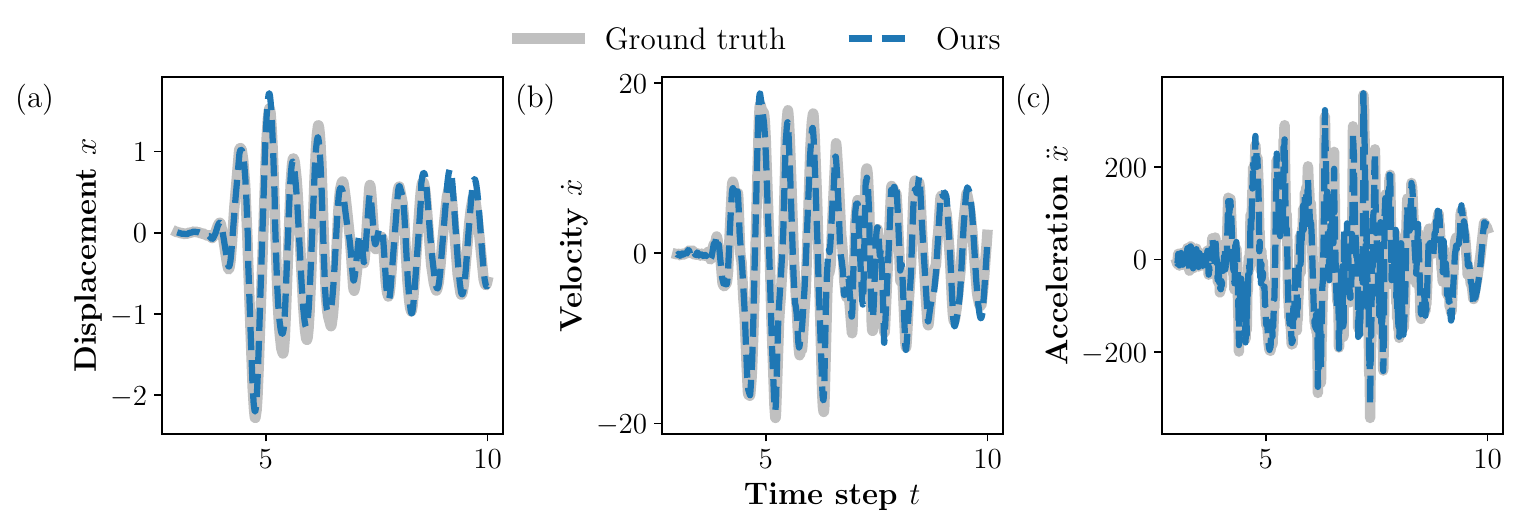}
    \caption{Testing results of SDOF experimental structure data without NSD. (a) Displacement. (b) Velocity. (c) Acceleration.}
    \label{fig:testing results without exp3}
\end{figure}

For the \textit{without NSD} case, the loop of $\dot{x}-x$ plane is presented in Figure \ref{fig:hysteresis loops exp3}(a). The discovered evolution law reduces to a linear form $\dot{z}(t) = 2.1847 \dot{x}(t)$, together with the motion equation $22.2798 \ddot{x}(t) + 18.4829 \dot{x}(t) + 1571.7874 x(t) + 1572.5618 z(t) = -22.2798 u(t)$, which in nature is a linear system. This reflects that our proposed framework is capable of automatically discovering a linear equation if the system comes without evident nonlinearity or hysteresis. Figure \ref{fig:training results without exp3} and \ref{fig:testing results without exp3} show that the model closely matches the training and testing responses in terms of displacement, velocity, and acceleration. Table \ref{tab:nrmse exp3} provides quantitative results. The small NRMSEs confirm that the framework achieves accurate and robust identification.

\begin{figure} [!h]
    \centering
    \includegraphics[width=1.0\linewidth]{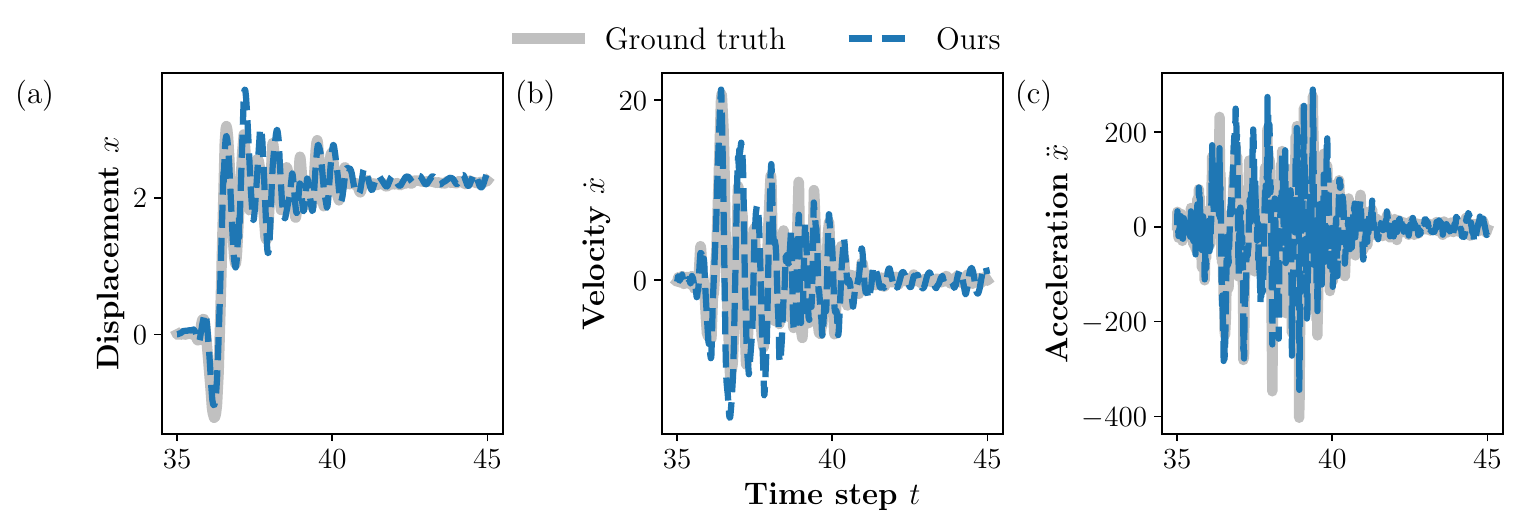}
    \caption{Training results of SDOF experimental structure data with NSD. (a) Displacement. (b) Velocity. (c) Acceleration.}
    \label{fig:training results with exp3}
\end{figure}

\begin{figure} [!t]
    \centering
    \includegraphics[width=1.0\linewidth]{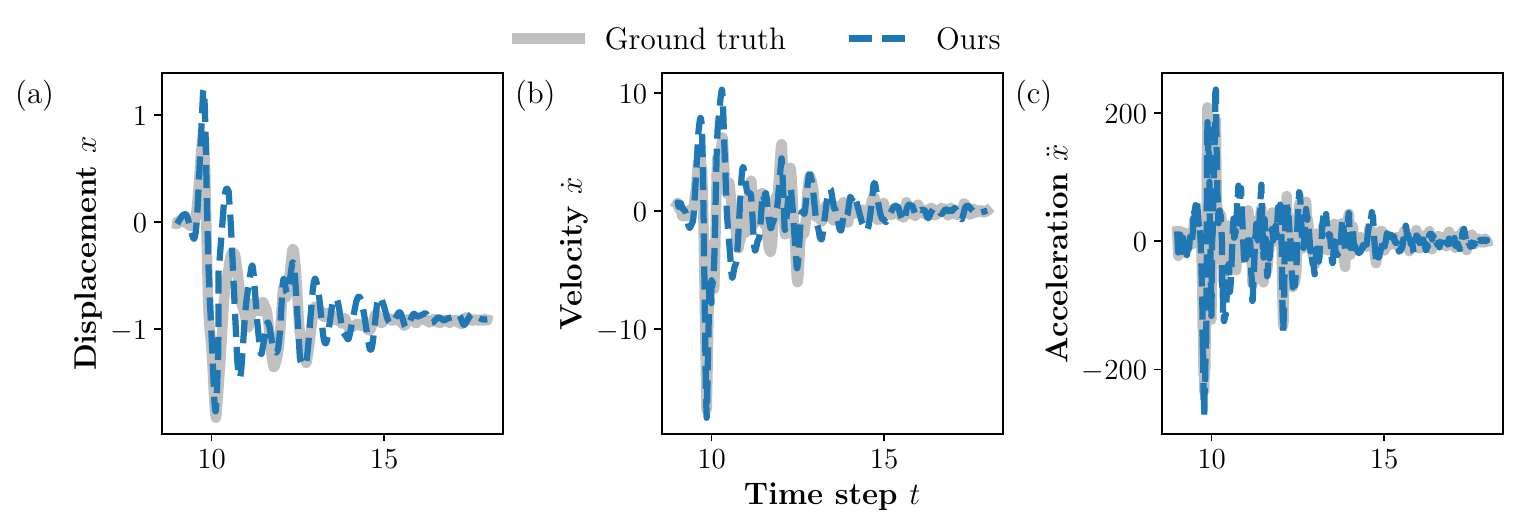}
    \caption{Testing results of SDOF experimental structure data with NSD. (a) Displacement. (b) Velocity. (c) Acceleration.}
    \label{fig:testing results with exp3}
\end{figure}

For the \textit{with NSD} case, the loops become much wider and exhibit obvious pinching and asymmetry (Figure \ref{fig:hysteresis loops exp3}(b)). Accordingly, the learned $\dot{z}(t)$ includes higher-order and mixed-magnitude terms of $x(t)$, $\dot{x}(t)$, and $z(t)$. The predicted hysteretic link equation is $\dot{z}(t) = 4.8586 \dot{x}(t) + 1.6412 x^2 |\dot{x}(t)|  + 1.8160 \dot{x}(t) |\dot{x}(t)| z(t) - 1.8174 \dot{x}(t) |z(t)| + 1.7648 z(t) |z(t)| $ with the dynamics motion equation $22.2798 \ddot{x}(t) + 44.7281 \dot{x}(t) + 2099.6526 x(t) + 2108.1516 z(t)$ $= -22.2798 u(t)$. 
As shown in Figure \ref{fig:training results with exp3} and \ref{fig:testing results with exp3}, the training responses are well captured, but testing displacement exhibits larger deviations. Table \ref{tab:nrmse exp3} confirms this: the displacement error rises to 14.11\% in the testing set, while velocity and acceleration remain within 6\%. 
This indicates that the NSD introduces stronger nonlinear hysteresis, which requires a richer $\dot{z}$ structure and leads to a larger generalization gap. 
\revised{The larger displacement error can be attributed to the stronger nonlinear hysteresis induced by the NSD, including wider hysteresis loops and asymmetry. In addition, the training and testing cases are driven by different earthquake records and therefore may cover different regions of the hysteretic response space (in the training data (Figure \ref{fig:training results with exp3}), the displacement is mostly positive, while in the test data (Figure \ref{fig:testing results with exp3}), the displacement is mostly negative). Low-frequency errors in the learned dynamics can accumulate more visibly in displacement, while velocity and acceleration remain less affected. Therefore, the current results should be interpreted as showing that the proposed framework captures the dominant nonlinear hysteretic trend in the \textit{with NSD case}, but does not fully eliminate the generalization gap under unseen strong nonlinear hysteresis.}

From all these results, it can be concluded that the proposed framework is capable of both equation discovery and dynamic prediction in real experimental data. 
For the relatively simple \textit{without NSD} case, it achieves concise and highly accurate identification, while for the more challenging \textit{with NSD} case, it still provides interpretable governing equations and meaningful forward predictions, albeit with a higher generalization error. 
In future work, these limitations may be further addressed by extending the framework towards more complex hysteretic systems, and improving its generalization to a wider range of nonlinear dynamical problems.

\begin{table}  [!h]
\centering
\caption{NRMSE results of SDOF experimental structure data.}
\resizebox{1.0\textwidth}{!}{%
\begin{tabular}{@{}c c c c c c @{}}
\toprule
\textbf{Scenarios} & \textbf{Responses} & 
\makecell[c]{\textbf{Training} \\ \textbf{NRMSE} \\ \textbf{(Ours)}}  & 
\makecell[c]{\textbf{Testing} \\ \textbf{NRMSE}  \\ \textbf{(Ours)}} &
\makecell[c]{\textbf{Training} \\ \textbf{NRMSE} \\ \textbf{(Sparse \cite{lai2019sparse})}}  & 
\makecell[c]{\textbf{Testing} \\ \textbf{NRMSE}  \\ \textbf{(Sparse \cite{lai2019sparse})}} \\
\midrule
\multirow{3}{*}{\makecell[c]{Without NSD}} 
& Displacement $x$         & 3.23\%   & 2.99\%   & 10.73\%   & 2.90\% \\ 
& Velocity $\dot{x}$       & 2.57\%   & 2.89\%   & 4.47\%   & 2.92\% \\
& Acceleration $\ddot{x}$  & 1.45\%   & 2.14\%   & 4.42\%   & 2.70\% \\
\midrule
\multirow{3}{*}{\makecell[c]{With NSD}} 
& Displacement $x$         & 4.12\%   & 14.11\%   & 6.57\%   & 13.85\% \\ 
& Velocity $\dot{x}$       & 4.49\%   & 5.60\%   & 4.56\%   & 5.62\% \\
& Acceleration $\ddot{x}$  & 5.96\%   & 4.86\%   & 5.71\%   & 6.26\% \\
\bottomrule
\end{tabular}
}
\label{tab:nrmse exp3}
\end{table}

\FloatBarrier

\begin{revisedblock}
\subsection{Extension to MDOF systems}
\label{subsec:mdof}

The proposed framework for equation discovery and dynamic prediction is not restricted to SDOF hysteretic systems. 
For a MDOF system~\cite{liu2025some,liu2024efficiency}, one can augment the state vectors, and represent the coupled dynamics in a matrix form ($t$ is omitted for brevity in this section) with $r$-DOF and
$q$ hysteretic variable(s).  
Therefore, the equations can be written as:
\begin{equation}
\left\{
\begin{aligned}
& \boldsymbol{M}\ddot{\boldsymbol{x}} + \boldsymbol{C}\dot{\boldsymbol{x}} + \boldsymbol{K}\boldsymbol{x} + \boldsymbol{H}\boldsymbol{z} = \boldsymbol{B}\boldsymbol{u}, \\
& \dot{\boldsymbol{z}} = \boldsymbol{g}_{\phi}\!\left(\dot{\boldsymbol{x}},\boldsymbol{z}\right),
\end{aligned}
\right.
\label{eq:mdof_coupled}
\end{equation}
where $\boldsymbol{x}\in\mathbb{R}^{r}$ denotes the displacement vector (with $r$-DOF), and $\boldsymbol{z}\in\mathbb{R}^{q}$ collects the hysteretic variables associated with localized hysteretic elements (e.g.,~inter-story links, yielding dampers, or absorber components). 
The matrices $\boldsymbol{M},\boldsymbol{C},\boldsymbol{K}\in\mathbb{R}^{r\times r}$ are the mass, damping, and stiffness matrices, respectively. The coupling matrix $\boldsymbol{H}\in\mathbb{R}^{r\times q}$ maps the hysteretic variables to generalized restoring forces, while $\boldsymbol{B}\in\mathbb{R}^{r\times p}$ distributes the external input $\boldsymbol{u}\in\mathbb{R}^{p}$ to the structural DOFs. The (possibly nonlinear) function $\boldsymbol{g}_{\phi}$ governs the hysteretic variable evolution and can encode rate dependence and memory effects.

To illustrate the above formulation, we consider a simple yet representative case with $r=2, q=1$ and $p=1$, corresponding to a 2-DOF model equipped with a single hysteretic energy-dissipating link. Let:
\begin{equation}
\boldsymbol{x}=
\begin{bmatrix}
x_{1}\\
x_{2}
\end{bmatrix}
,
\qquad
z\in\mathbb{R},
\label{eq:2dof_state_def}
\end{equation}
where $x_{1}$ and $x_{2}$ are the displacements of the 2-DOFs, and $z$ represents the hysteretic variable. The coupled dynamics in Eq.~\eqref{eq:mdof_coupled} can be reduce to:
\begin{equation}
\left\{
\begin{aligned}
& \boldsymbol{M}\ddot{\boldsymbol{x}} + \boldsymbol{C}\dot{\boldsymbol{x}} + \boldsymbol{K}\boldsymbol{x} + \boldsymbol{H}z = \boldsymbol{B}u, \\
& \dot{z} = g_{\phi}\!\left(\dot{x}_1,\dot{x}_2,z\right).
\end{aligned}
\right.
\label{eq:2dof_1int_coupled}
\end{equation}

As a concrete numerical example, we choose a harmonic excitation $u(t)=1000 \sin\!\left(2\pi \times 1.5 t \right)$ and $\mathbf{B}=[1,1]^T$, specify:
\begin{equation}
\boldsymbol{M}=
\begin{bmatrix}
m_1 & 0\\
0 & m_2
\end{bmatrix}
=
\begin{bmatrix}
100 & 0\\
0 & 80
\end{bmatrix}
,
\quad
\boldsymbol{C}=
\begin{bmatrix}
c_{11} & c_{12}\\
c_{21} & c_{22}
\end{bmatrix}
=
\begin{bmatrix}
30 & -5\\
-5 & 20
\end{bmatrix}
,
\label{eq:2dof_mck}
\end{equation}
\begin{equation}
\boldsymbol{K}=
\begin{bmatrix}
k_{11} & k_{12}\\
k_{21} & k_{22}
\end{bmatrix}
=
\begin{bmatrix}
3500 & -1500\\
-1500 & 1500
\end{bmatrix}
,
\quad
\boldsymbol{H}=
\begin{bmatrix}
h_1\\
h_2
\end{bmatrix}
=
\begin{bmatrix}
1\\
-1
\end{bmatrix}
.
\label{eq:2dof_hb}
\end{equation}

With this choice, the hysteretic link exerts equal and opposite restoring forces on the 2-DOFs. For the hysteretic variable evolution, we adopt a simple rate-dependent form driven by the inter-story relative velocity $\dot{x}_2-\dot{x}_1$:
\begin{equation}
\dot{z} = a(\dot{x}_2-\dot{x}_1) - b\left|\dot{x}_2-\dot{x}_1\right|\,z,
\qquad
a=50,\;\; b=2.
\label{eq:2dof_zdot}
\end{equation}

For completeness, Eq.~\eqref{eq:2dof_1int_coupled} can be written component-wise as:
\begin{equation}
\label{eq:2dof_componentwise}
\left\{
\begin{aligned}
& 100\,\ddot{x}_1 + 30\,\dot{x}_{1} - 5\,\dot{x}_2 + 3500\,x_1 - 1500\,x_2 + z = u,\\
& 80\,\ddot{x}_2 - 5\,\dot{x}_{1} + 20\,\dot{x}_2 - 1500\,x_1 + 1500\,x_2 - z = u,\\
& \dot{z} = 50(\dot{x}_2-\dot{x}_1) - 2\left|\dot{x}_2-\dot{x}_1\right|\,z,
\end{aligned}
\right.
\end{equation}
which represent the dynamic motion equation $f_{\theta_1}, f_{\theta_2}$ and the hysteretic link equation $g_\phi$, respectively.

The above derivations illustrate that the adopted state-space formulations extend naturally to MDOF settings by treating both structural states and hysteretic variables in a coupled form.
Therefore, the formulation remains structurally identical to the SDOF case, with the equation discovery task now applied to Eq.~\eqref{eq:2dof_componentwise}. In what follows, we present the \textit{Hysteresis Discovery} case in Figure \ref{fig:structure}, where the measurements are noise-free and $\mathbf{x}$, $\dot{\mathbf{x}}$, and $\ddot{\mathbf{x}}$ are all known, and the other settings are similar. Combining the procedure in the proposed framework, the unknowns include the entries of $\mathbf{M}$, $\mathbf{C}$, $\mathbf{K}$, and $\mathbf{H}$, as well as the entire functional form of the hysteretic-variable evolution in Eq.~\eqref{eq:2dof_zdot}. 
Using the strategy of reformulation introduced in Section \ref{sec3}, the MDOF formulation can be written explicitly in a first-order state-space form:
\begin{equation}
\frac{d}{dt}
\begin{bmatrix}
x_1 \\
x_2 \\
\dot{x}_1 \\
\dot{x}_2
\end{bmatrix}
=
\begin{bmatrix}
\dot{x}_1 \\
\dot{x}_2 \\
f_{\theta_1}(x_1, x_2, \dot{x}_1, \dot{x}_2, u, z) \\
f_{\theta_2}(x_1, x_2, \dot{x}_1, \dot{x}_2, u, z)
\end{bmatrix}
,
\end{equation}
with $z$ being trainable.   

The discovered governing equations and NRMSE results can be seen in Table \ref{tab:mdof_eq} and \ref{tab:mdof_nrmse}.
Figure \ref{fig:exp4} shows the training and testing results of the discovered MDOF system for both displacement and velocity responses.
These results indicate that the proposed framework remains effective when extended from SDOF to MDOF systems, yielding reasonably accurate equation discovery and dynamic predictions. 

We acknowledge that, as in many MDOF settings, the problem becomes increasingly complex as the number of DOF grows. As $r$ and $q$ increase, both computational cost and identifiability challenges can become more pronounced, especially when hysteresis is present in multiple or adjacent DOFs due to structural coupling. 
\revised{The purpose of this proof-of-concept example is to demonstrate the applicability of the framework to a coupled MDOF setting, while its scalability to general high-dimensional systems, especially with multiple coupled hysteretic variables, should be further examined in future work.}
Nevertheless, based on above observations, the proposed framework remains directly applicable and can handle this increased complexity.

\begin{table}[!h]
\centering
\captionsetup{labelfont={color=black}, textfont={color=black}}
{\revised{%
\caption{Discovered governing equations of MDOF hysteretic system.}
\label{tab:mdof_eq}
\resizebox{1.0\textwidth}{!}{%
\begin{tabular}{c l}
\toprule
\textbf{Type} & \makecell[c]{\textbf{Equation}} \\
\midrule
True $f_{\theta_1}$  &
$100.0000\,\ddot{x}_1 + 30.0000\,\dot{x}_{1} - 5.0000\,\dot{x}_2 + 3500.0000\,x_1 - 1500\,x_2 + 1.0000z = u$ \\
\midrule
Discovered $f_{\theta_1}$  &
$94.9694 \ddot{x}_{1} + 31.5296 \dot{x}_{1} - 4.7204 \dot{x}_{2} + 3325.0000 x_{1} - 1574.9634 x_{2} + 0.9799 z = u$ \\
\midrule
True $f_{\theta_2}$ &
$80.0000\,\ddot{x}_2 - 5.0000\,\dot{x}_{1} + 20.0000\,\dot{x}_2 - 1500.0000\,x_1 + 1500\,x_2 - 1.0000z = u$ \\
\midrule
Discovered $f_{\theta_2}$ &
$83.9695 \ddot{x}_{2} - 5.2208 \dot{x}_{1} + 19.0294 \dot{x}_{2} - 1424.9634 x_{1} + 1575.0122 x_{2} - 1.0200 z = u$ \\
\midrule
True $g_{\phi}$ &
$\dot{z} = 50.0000(\dot{x}_2-\dot{x}_1) - 2.0000\left|\dot{x}_2-\dot{x}_1\right|\,z$ \\
\midrule
Discovered $g_{\phi}$ &
$\dot{z} = 46.1250(\dot{x}_2 - \dot{x}_1) - 2.3642 \left|\dot{x}_2-\dot{x}_1\right|\,z$ \\
\bottomrule
\end{tabular}%
}
}}
\end{table}

\begin{table}[!h]
\centering
\captionsetup{labelfont={color=black}, textfont={color=black}}
{\revised{%
\caption{NRMSE results of MDOF hysteretic system.}
\resizebox{0.75\textwidth}{!}{%
\label{tab:mdof_nrmse}
\begin{tabular}{c c c}
\toprule
\textbf{Responses} & \textbf{Training NRMSE} & \textbf{Testing NRMSE} \\
\midrule
Displacement $x_1$       & 2.55\% & 5.99\% \\
Displacement $x_2$       & 1.35\% & 3.73\% \\
Velocity $\dot{x}_1$ & 3.44\% & 4.97\% \\
Velocity $\dot{x}_2$ & 2.95\% & 2.09\% \\
\bottomrule
\end{tabular}
}
}}
\end{table}

\begin{figure} [!h]
    \centering
    \captionsetup{labelfont={color=black}, textfont={color=black}}
    \includegraphics[width=1.0\linewidth]{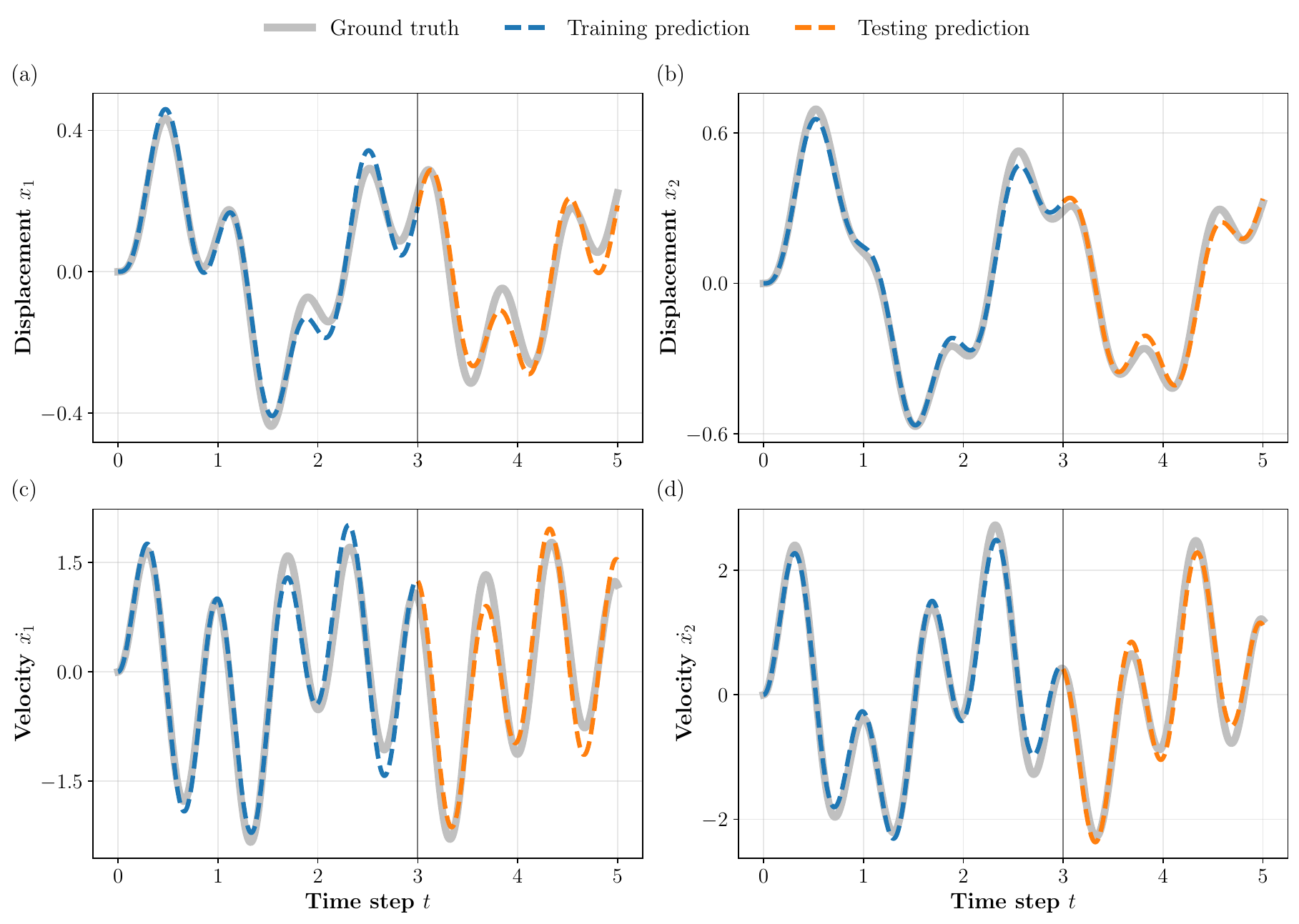}
    \caption{Training and testing results of MDOF hysteretic system. The time interval $0$--$3~\mathrm{s}$ is used for training, and $3$--$5~\mathrm{s}$ is used for testing. (a)\&(b) Displacement results. (c)\&(d) Velocity results.}
    \label{fig:exp4}
\end{figure}

\FloatBarrier
\end{revisedblock}

\clearpage
\section{Conclusion}  \label{sec5}
This research presented a unified framework for equation discovery and dynamic prediction of hysteretic systems, with a proposed solver-based hysteretic variable learning scheme that addresses one of the key challenges in hysteresis modeling -- the difficulty of inferring hysteretic variables.
We summarize existing work on equation discovery for hysteretic systems and categorize them according to the level of assumed available knowledge. 
The proposed unified framework in our work is capable of addressing two major challenges of hysteretic systems across various scenarios: the learning of unobservable hysteretic variables and the discovery of governing equations with complex nonlinear expressions. 
The effectiveness of the framework is validated through \revised{four} representative case studies. 
On the Bouc-Wen benchmark, it accurately recovers the governing equations even in the presence of noise, confirming its robustness and reliability. 
For a more complex nonlinear structure with cubic stiffness and fractional hysteretic exponents, the framework successfully achieves superior accuracy and interpretability compared with library-based methods such as SINDy \revised{and its variations}. 
In the real shake-table experiment of a yielding structure, the framework reveals concise linear dynamics without NSD and more complex nonlinear hysteresis behavior with NSD, achieving stable predictions.
\revised{The extension to MDOF systems provides a preliminary proof-of-concept showing that the proposed framework can be applied to a coupled MDOF system.}

This research not only provides insights into equation discovery and dynamic prediction of hysteretic systems but also demonstrates the potential of the proposed framework for broader applications in nonlinear dynamical systems, including those with or without hysteretic variables. 
Its flexibility and interpretability make it applicable to various domains such as structural dynamics, material modeling, and control system identification, where nonlinear and memory-dependent behaviors are common.  
\revised{Although the framework has extended the validation to fractional and MDOF settings, further investigation of additional hysteresis families, such as strongly asymmetric, severely pinched, S-shaped, or flag-shaped loops, remains important in the future.}
Future work will focus on enhancing the efficiency and scalability of the framework, improving its ability to handle more complex systems, and exploring probabilistic or hybrid modeling strategies to further extend its applicability to real-world engineering and scientific problems.

\newpage
\section*{Acknowledgments}
The authors would like to express their sincere gratitude to Prof.~Satish Nagarajaiah at Rice University for generously sharing the SDOF experimental structure data that made this research possible. 
The authors wish to express their gratitude for the financial support received from {Guangdong Provincial Fund - Special Innovation Project} ({2024KTSCX038}); {Research Grants Council of Hong Kong through the Research Impact Fund} ({R5006-23}); {Guangdong Provincial Key Lab of Integrated Communication, Sensing} and {Computation for Ubiquitous Internet of Things} (No. {2023B1212010007}); {Natural Science Foundation of Guangdong Province, China} (General Fund; Grant No. {2026A1515011629}).

\begin{revisedblock}
\appendix
\section{Network architecture and training details}
\label{Network architecture}
In this work implementation, $\mathcal{N}_\theta(\cdot)$ in \textit{Full Equation Discovery} is parameterized as a multilayer perceptron with two hidden layers, each of width 128, and hyperbolic tangent activation functions. The network parameters are optimized jointly within the solver-based training procedure using the Adam algorithm, with a learning rate of $3\times10^{-4}$, first- and second-moment decay rates of 0.9 and 0.98, respectively, and an $\ell_2$ regularization coefficient of $10^{-4}$. To improve optimization stability, gradient norm clipping with a threshold of 1.0 is applied. Training is terminated either upon convergence of the loss or when a prescribed maximum number of iterations (e.g.,~10,000) is reached.

\section{Computational complexity and scalability}
\label{sec:complexity}
We briefly discuss computational complexity and scalability as the system dimension increases. Let $r$ denote the number of degrees of freedom (DOFs), $q$ the number of hysteretic variables, and $N$ the total number of sampled time points.

\subsection{Step 1: Solver-based hysteretic variable learning.}
At this step, the trainable hysteretic variable is learned by embedding the governing dynamics into a differentiable ODE solver and jointly optimizing the model parameters and the trainable hysteretic variables through trajectory matching.
Since the trainable hysteretic variable is parameterized outside the solver (shown in  Figure \ref{fig:framework}), the embedded solver advances only the displacement and velocity states, so the solver state has dimension $2r$.
For each optimization iteration, the solver is unrolled over $N$ time samples, and the trainable hysteretic variable is injected into the motion equation at the corresponding sampled times.
If the solver uses $n_{\mathrm{s}}$ stage evaluations per time step, with $n_{\mathrm{s}}=4$ for RK4, the per-iteration solver cost is $\mathcal{O}(N n_{\mathrm{s}}\, 2r)$, while updating the trainable hysteretic variables over all time samples introduces an additional cost of $\mathcal{O}(N q)$ per iteration.
Hence, the per-iteration computational cost is $\mathcal{O}\!\left(N n_{\mathrm{s}}\, 2r + N q\right)$.
Since $n_{\mathrm{s}}$ is typically small and fixed in practice, this can be simplified to:
\begin{equation}
\mathcal{O}\!\left(N(2r+q)\right).
\end{equation}

\subsection{Step 2: Equation discovery.}
At this step, SR is applied to the learned trajectories to recover explicit governing equations.
The computational cost is dominated by the symbolic search process.
Let $P$ denote the population size and $G$ the number of generations.
For each candidate symbolic expression, the main cost comes from evaluating the expression over the available data samples.
If $C$ denotes the per-sample cost of evaluating one candidate symbolic expression, then the computational cost of Step 2 is:
\begin{equation}
\mathcal{O}\!\left(N\,P\,G\,C\right).
\end{equation}

Relative to conventional gradient-based parameter estimation for a prescribed hysteretic model, the proposed Step 1 shares a similar computational backbone, as both require repeated forward ODE integration and gradient-based optimization over trajectory data. 
However, the present framework is generally more computationally demanding in practice because the hysteretic relation is not fixed a priori and must instead be learned in a more flexible function space, rather than calibrated through a small set of prescribed physical parameters. 
In addition, the proposed framework introduces an offline Step 2 after training. 
These additional computational costs represent a trade-off for reduced structural assumptions and the ability to recover an interpretable symbolic governing equation.

\subsection{Impact of multiple hysteretic variables.} 
When hysteresis is present at multiple or adjacent DOFs, different hysteretic variables may contribute to overlapping restoring-force components because of structural coupling. 
As a result, similar observable responses can be explained by different combinations of hysteretic variables, which makes the different hysteretic contributions harder to identify and renders the joint optimization more ill-conditioned. 
From both computational and identifiability perspectives, scalability therefore relies not only on computational resources, but also on sufficiently informative excitations, multiple training trajectories, adequate sensing, and physically motivated regularization (e.g.,~sparsity or locality in cross-DOF coupling) to stabilize learning and improve the separability of hysteretic contributions.

\section{Additional experiments on the Bouc-Wen hysteretic system benchmark}
\label{bouc add}
To further validate the effectiveness and practical applicability of the proposed framework, we conduct some additional benchmark experiments beyond the baseline setting.

\subsection{Cross-excitation validation} 
To assess whether the discovered equations are biased to a specific excitation history, we swapped the training and testing roles of the two excitation signals used in the baseline in the medium-noise setting (SNR = 20~dB). Specifically, the multisine signal is used for training and the sinesweep signal for testing, i.e.,~the reverse of the setting shown in Figure \ref{fig:excitation exp1}.
For a fair comparison, all remaining training settings (data length, sampling rate, optimization hyperparameters, etc.) are kept identical to the baseline.

For quantitative evaluation, Table \ref{tab:cross&extra exp1} reports the NRMSE results, together with discovered equations $1.7803 \ddot{x} + 8.0472 \dot{x} + 47360.7258 x + 0.8736 z = u$ and $\dot{z} = 57072.5992 \dot{x} - 748.504 |\dot{x}|  z + 748.504 \dot{x} |z| $.
Figure \ref{fig:displacement results exp1 cross} and \ref{fig:velocity results exp1 cross} visualize the predicted responses against the ground truth in the swapped excitation setting.
In this cross-excitation validation example, the proposed framework maintains competitive accuracy for both seen and unseen excitation types, suggesting that the discovered equations are not related to a specific input history and retain predictive capability across the two benchmark excitation types.


\begin{table} [!h]
\centering
\captionsetup{labelfont={color=black}, textfont={color=black}}
{\revised{%
\caption{NRMSE results for the cross-excitation validation and extrapolation tests on the benchmark system.}
\resizebox{0.95\textwidth}{!}{
\begin{tabular}{@{}c c c c c@{}}
\toprule
\textbf{Category} & \textbf{Responses} & 
\makecell[c]{\textbf{Training} \\ \textbf{NRMSE}}  & 
\makecell[c]{\textbf{Testing 1} \\ \textbf{NRMSE}} & 
\makecell[c]{\textbf{Testing 2} \\ \textbf{NRMSE}} \\
\midrule
\multirow{2}{*}{Baseline} 
& Displacement $x$   & 2.41\%  & 1.64\%  & 3.30\% \\
& Velocity $\dot{x}$ & 3.08\%  & 2.15\%  & 3.37\% \\
\midrule
\multirow{2}{*}{Cross-excitation validation} 
& Displacement $x$   & 2.71\%  & 3.03\%  & 3.20\% \\
& Velocity $\dot{x}$ & 3.11\%  & 3.13\%  & 2.87\% \\
\midrule
\multirow{2}{*}{Extrapolation test} 
& Displacement $x$   & --       & 1.93\%  & 3.24\% \\
& Velocity $\dot{x}$ & --       & 2.63\%  & 3.77\% \\
\bottomrule
\end{tabular}
}
\label{tab:cross&extra exp1}
}}
\end{table}

\begin{figure} [!h]
    \centering
    \captionsetup{labelfont={color=black}, textfont={color=black}}
    \includegraphics[width=1.0\linewidth]{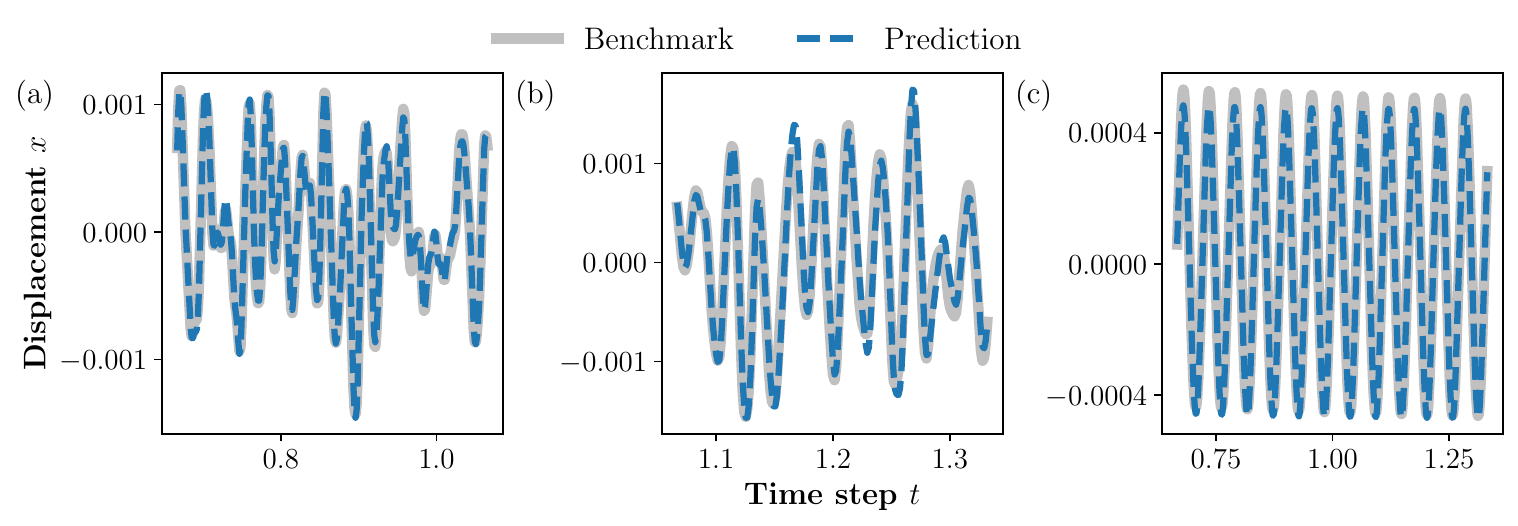}
    \caption{Displacement results on the benchmark system - cross-excitation validation. In contrast to Figure \ref{fig:displacement results exp1}, the multisine signal is used as the training input, while the sinesweep signal is used as the testing input. (a) Training results. (b) Testing 1 (same excitation) results. (c) Testing 2 (unseen excitation) results.}
    \label{fig:displacement results exp1 cross}
\end{figure}

\begin{figure} [!h]
    \centering
    \captionsetup{labelfont={color=black}, textfont={color=black}}
    \includegraphics[width=1.0\linewidth]{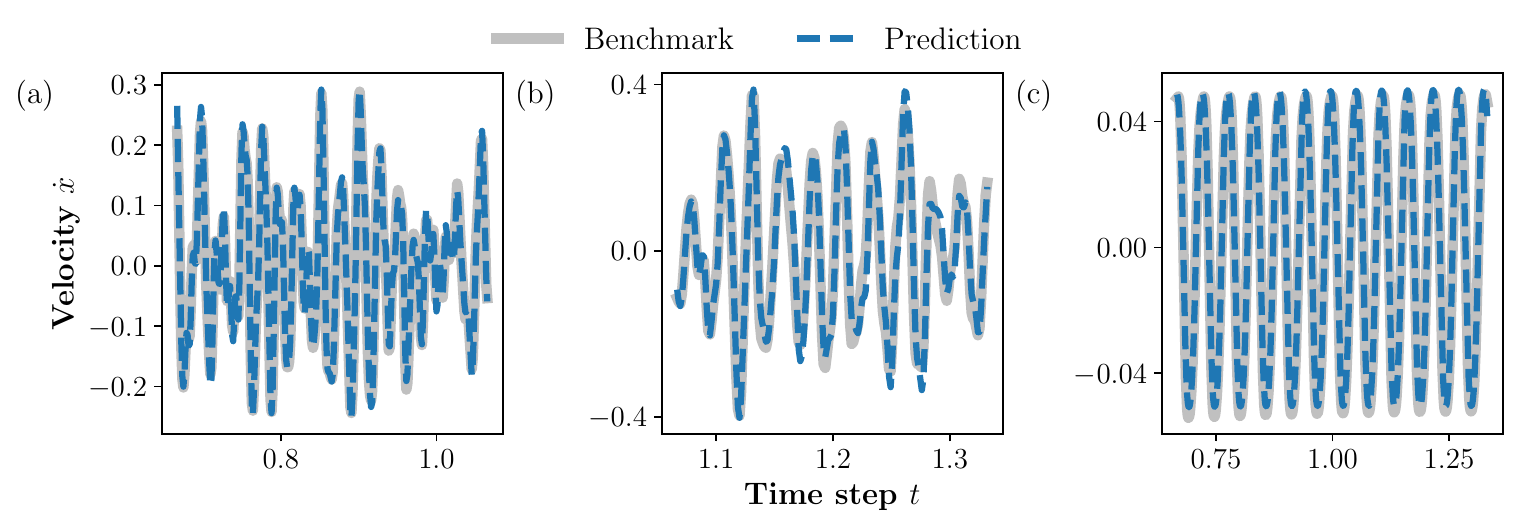}
    \caption{Velocity results on the benchmark system - cross-excitation validation. In contrast to Figure \ref{fig:velocity results exp1}, the multisine signal is used as the training input, while the sinesweep signal is used as the testing input. (a) Training results. (b) Testing 1 (same excitation) results. (c) Testing 2 (unseen excitation) results.}
    \label{fig:velocity results exp1 cross}
\end{figure}

\subsection{Extrapolation test on unseen amplified excitation} 
To evaluate performance beyond the training operating range, we keep the same medium-noise baseline setting and test the discovered equations under an amplitude-scaled excitation without retraining.
In contrast to the nominal testing condition, we construct an amplitude-extrapolated test input by scaling the multisine excitation by a factor of 1.5, while keeping all other settings unchanged.
Without retraining or re-identification, we evaluate the governing equations obtained under the baseline setting.

The quantitative NRMSE results are summarized in Table \ref{tab:cross&extra exp1}, and the corresponding displacement and velocity comparisons are visualized in Figure \ref{fig:displacement results exp1 extra} and \ref{fig:velocity results exp1 extra}.
From this example, the proposed framework remains stable under the increased excitation amplitude and achieves accuracy comparable to that under the nominal testing condition, indicating reasonable extrapolation capability within the considered amplitude range.

\begin{figure} [!h]
    \centering
    \captionsetup{labelfont={color=black}, textfont={color=black}}
    \includegraphics[width=1.0\linewidth]{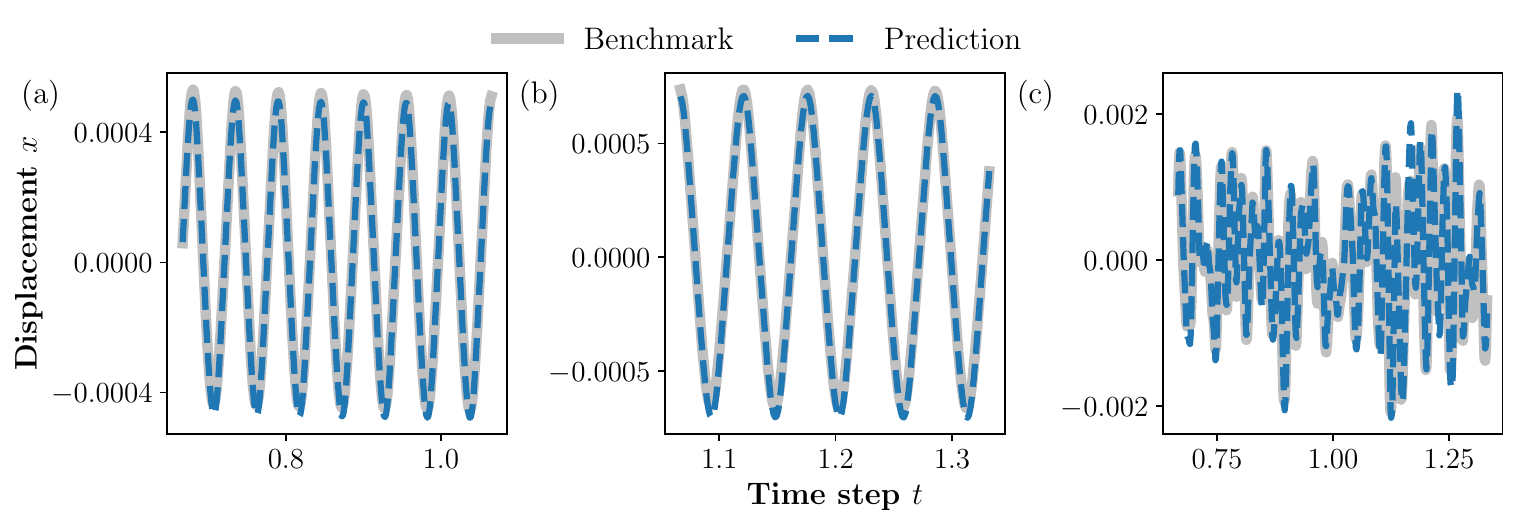}
    \caption{Displacement results on the benchmark system - extrapolation test. In contrast to  Figure \ref{fig:displacement results exp1}, the multisine excitation used for testing is amplitude-scaled by a factor of 1.5. (a) Training results. (b) Testing 1 (same excitation) results. (c) Testing 2 (unseen excitation) results.}
    \label{fig:displacement results exp1 extra}
\end{figure}

\clearpage

\begin{figure} [!h]
    \centering
    \captionsetup{labelfont={color=black}, textfont={color=black}}
    \includegraphics[width=1.0\linewidth]{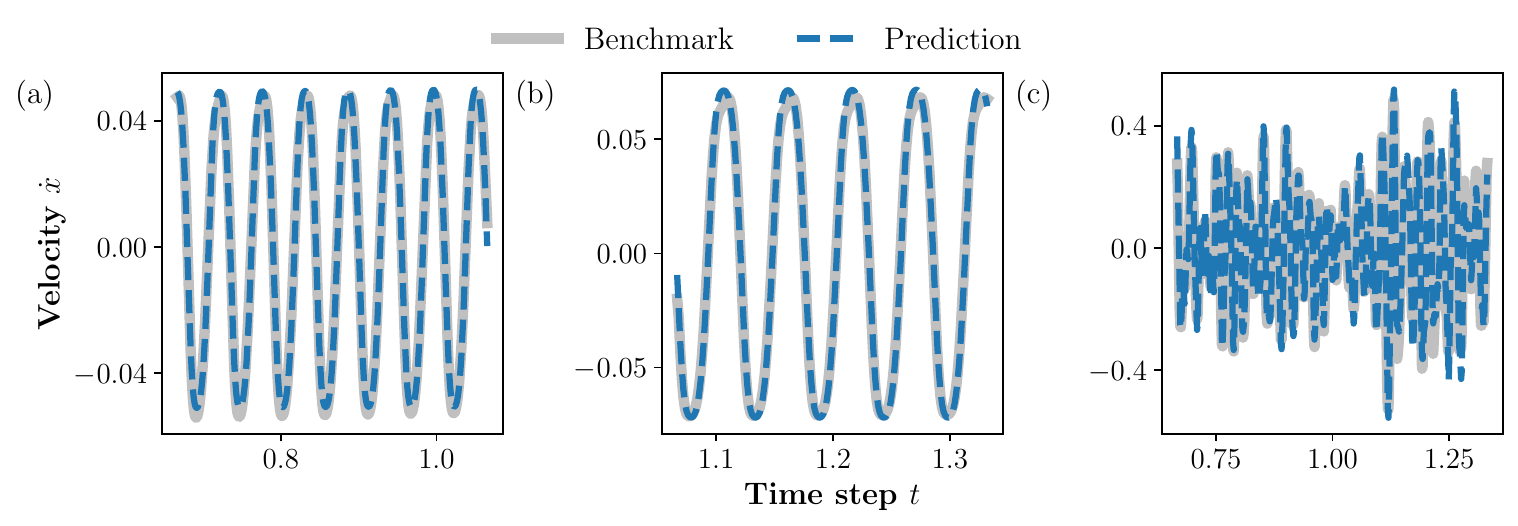}
    \caption{Velocity results on the benchmark system - extrapolation test. In contrast to  Figure \ref{fig:velocity results exp1}, the multisine excitation used for testing is amplitude-scaled by a factor of 1.5. (a) Training results. (b) Testing 1 (same excitation) results. (c) Testing 2 (unseen excitation) results.}
    \label{fig:velocity results exp1 extra}
\end{figure}

\subsection{Noisy wideband random excitation training}
To assess whether the proposed framework is suitable for noisy and new excitation, a band-limited wideband random excitation is synthesized in the frequency domain and transformed back to the time domain, after which measurement noise is added to emulate a noisy input signal. The same medium-noise setting is adopted for both the system responses and the excitation measurement. 

Table \ref{tab:robust_extension exp1} reports the NRMSE results. In this experiment, the model is trained with an excitation signal different from that in the baseline, while the testing excitation is kept unchanged for a fair comparison.  Figure \ref{fig:displacement results exp1 noisy wideband} and \ref{fig:velocity results exp1 noisy wideband} visualize the predicted displacement and velocity responses. The results show that the proposed framework maintains good predictive accuracy under the new and noisy excitation condition.

\begin{table} [!h]
\centering
\captionsetup{labelfont={color=black}, textfont={color=black}}
{\revised{%
\caption{NRMSE results for the robustness extension experiments on the benchmark system.}
\resizebox{0.95\textwidth}{!}{
\begin{tabular}{@{}c c c c c@{}}
\toprule
\textbf{Category} & \textbf{Responses} & 
\makecell[c]{\textbf{Training} \\ \textbf{NRMSE}}  & 
\makecell[c]{\textbf{Testing 1} \\ \textbf{NRMSE}} & 
\makecell[c]{\textbf{Testing 2} \\ \textbf{NRMSE}} \\
\midrule
\multirow{2}{*}{Baseline} 
& Displacement $x$   & 2.41\%  & 1.64\%  & 3.30\% \\
& Velocity $\dot{x}$ & 3.08\%  & 2.15\%  & 3.37\% \\
\midrule
\multirow{2}{*}{\makecell[c]{Noisy wideband random \\ excitation training}} 
& Displacement $x$   & 2.18\%  & 2.94\%  & 3.11\% \\
& Velocity $\dot{x}$ & 2.17\%  & 3.00\%  & 2.23\% \\
\midrule
\multirow{2}{*}{Without hysteresis} 
& Displacement $x$   & 0.88\%  & 0.87\%  & 0.82\% \\
& Velocity $\dot{x}$ & 0.61\%  & 1.04\%  & 0.77\% \\
\bottomrule
\end{tabular}
}
\label{tab:robust_extension exp1}
}}
\end{table}

\begin{figure} [!h]
    \centering
    \captionsetup{labelfont={color=black}, textfont={color=black}}
    \includegraphics[width=1.0\linewidth]{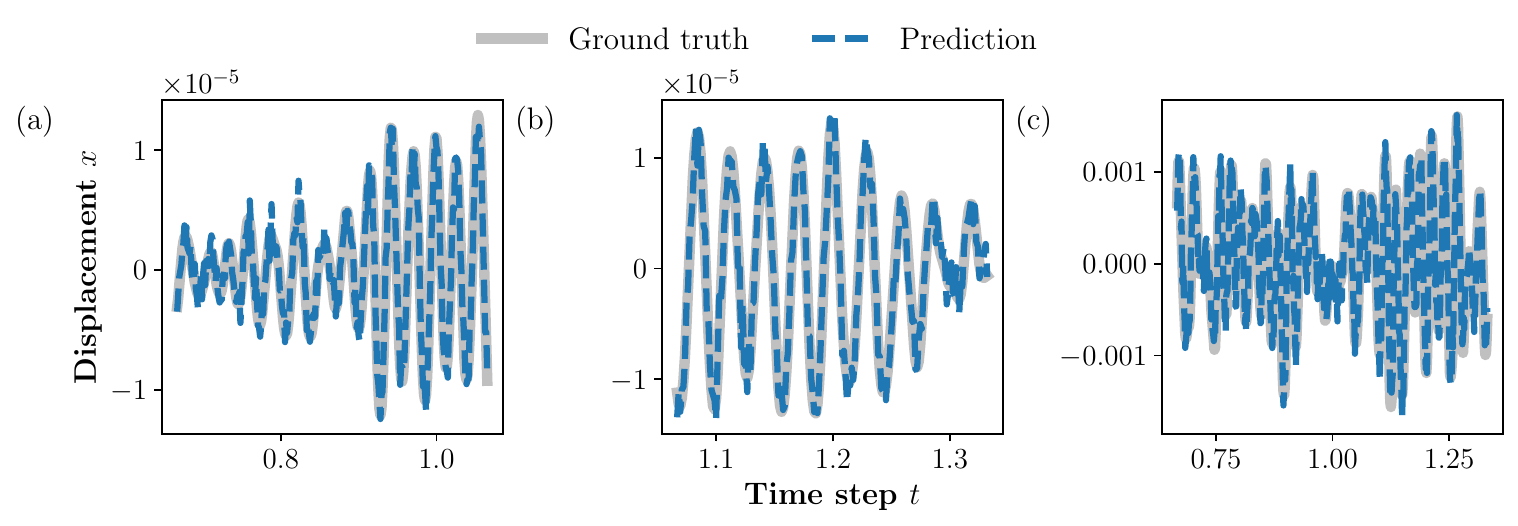}
    \caption{Displacement results on the benchmark system - noisy wideband random excitation training. (a) Training results. (b) Testing 1 (same excitation) results. (c) Testing 2 (unseen excitation) results.}
    \label{fig:displacement results exp1 noisy wideband}
\end{figure}

\begin{figure} [!h]
    \centering
    \captionsetup{labelfont={color=black}, textfont={color=black}}
    \includegraphics[width=1.0\linewidth]{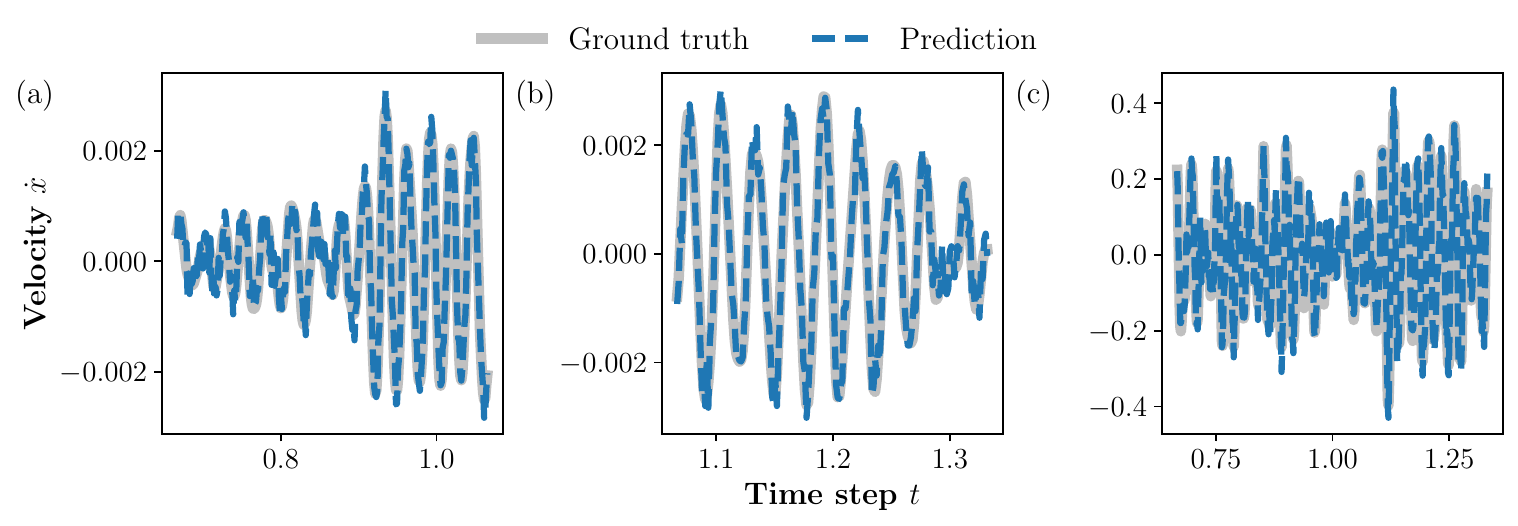}
    \caption{Velocity results on the benchmark system - noisy wideband random excitation training. (a) Training results. (b) Testing 1 (same excitation) results. (c) Testing 2 (unseen excitation) results.}
    \label{fig:velocity results exp1 noisy wideband}
\end{figure}

\subsection{Without hysteresis}
To further assess practical applicability, we test whether the framework reduces to an appropriate non-hysteretic model when hysteresis is absent.
We construct a non-hysteretic benchmark by setting $\beta=0$ and $\gamma=0$ in Eq.~\eqref{bouc-wen benchmark eq}, so that the hysteretic link equation reduces to $\dot{z}(t)=A\dot{x}(t)$. Neglecting the integration constant, we obtain $z(t)=Ax(t)$. Therefore, the true Eq.~\eqref{bouc-wen benchmark eq} is $2 \ddot{x}(t) + 10 \dot{x}(t) + 100000 x(t) = u(t)$, degenerate to a non-hysteretic case. All other training and evaluation settings are kept identical to baseline.

In this without hysteresis experiment, the discovered dynamic motion equation is $1.9998\ddot{x}(t) + 10.0049 \dot{x}(t) + 49999.4985 x(t) + 1.0089 z(t) = u(t)$, and the discovered hysteretic link equation is $\dot{z}(t) = 49999.5616 \dot{x}(t)$. Substituting the hysteretic link relation into the motion equation yields the degenerate non-hysteretic form $1.9998\ddot{x}(t) + 10.0049 \dot{x}(t) + 100444.0562 x(t) = u(t)$, which is consistent with the expected non-hysteretic reduction. 
Figure \ref{fig:hysteresis loops exp1 without hysteresis} shows the hysteresis loops. In panel (c), $F = kx + \alpha z = (k+A)x$, and panels (b) and (c) clearly exhibit an approximately linear relationship in the degenerate regime, consistent with the above analysis. The NRMSE results in Table \ref{tab:robust_extension exp1} further confirm that the proposed framework remains accurate when the dynamics reduce to the simpler non-hysteretic case. This is also supported by the response comparisons in Figure \ref{fig:displacement results exp1 without hysteresis} and \ref{fig:velocity results exp1 without hysteresis}. 
These results further support the unified nature of the proposed framework, which can recover both hysteretic and degenerate non-hysteretic dynamics without requiring a prescribed model structure.

\begin{figure} [!h]
    \centering
    \captionsetup{labelfont={color=black}, textfont={color=black}}
    \includegraphics[width=1.0\linewidth]{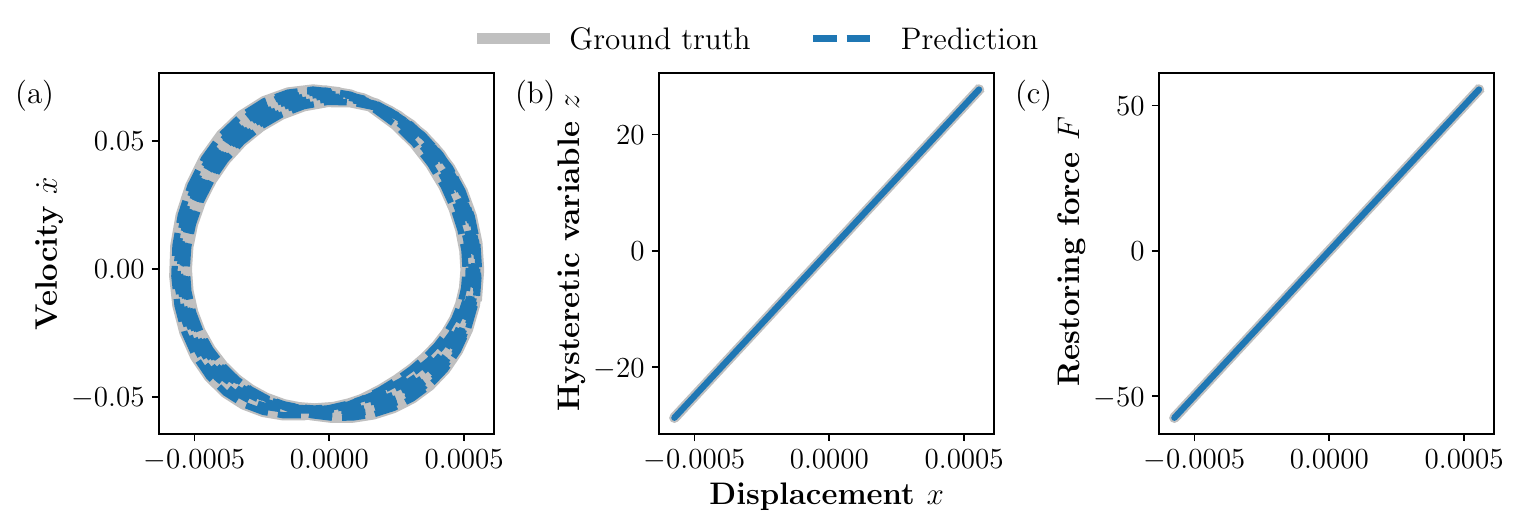}
    \caption{Hysteresis loops on the benchmark system - without hysteresis. (a) $x-\dot{x}$ hysteresis loop. (b) $x-z$ hysteresis loop. (c) $x-F (F = (k+A)x)$ hysteresis loop.}
    \label{fig:hysteresis loops exp1 without hysteresis}
\end{figure}

\begin{figure} [!h]
    \centering
    \captionsetup{labelfont={color=black}, textfont={color=black}}
    \includegraphics[width=1.0\linewidth]{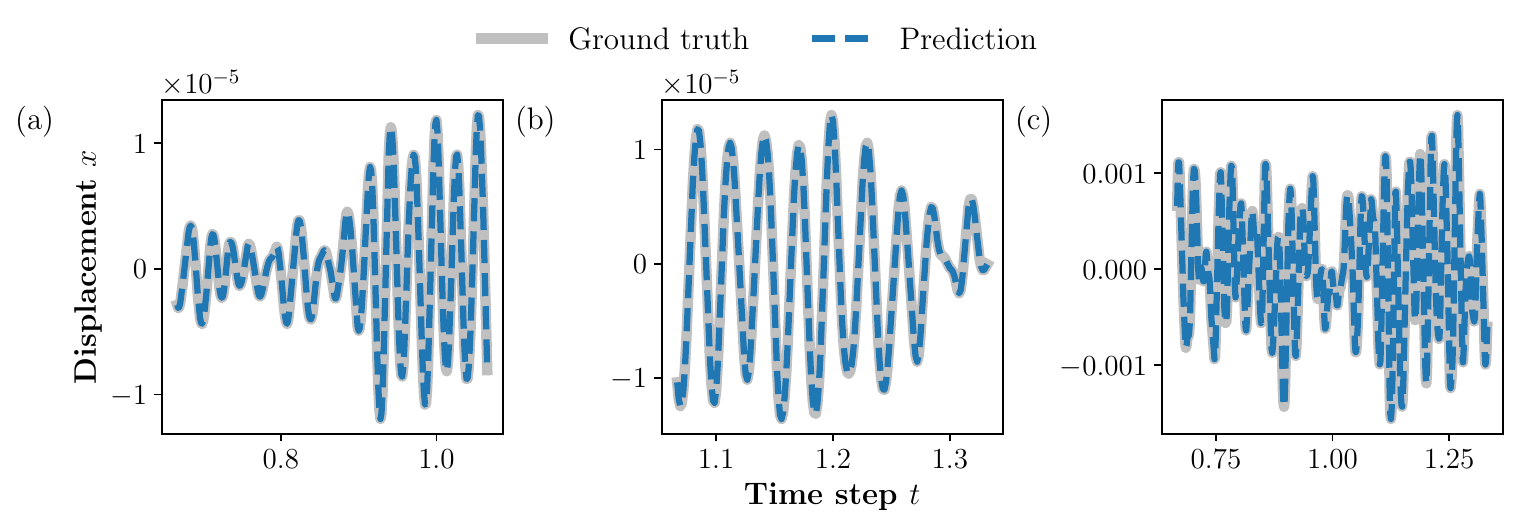}
    \caption{Displacement results on the benchmark system - without hysteresis. (a) Training results. (b) Testing 1 (same excitation) results. (c) Testing 2 (unseen excitation) results.}
    \label{fig:displacement results exp1 without hysteresis}
\end{figure}

\begin{figure} [!h]
    \centering
    \captionsetup{labelfont={color=black}, textfont={color=black}}
    \includegraphics[width=1.0\linewidth]{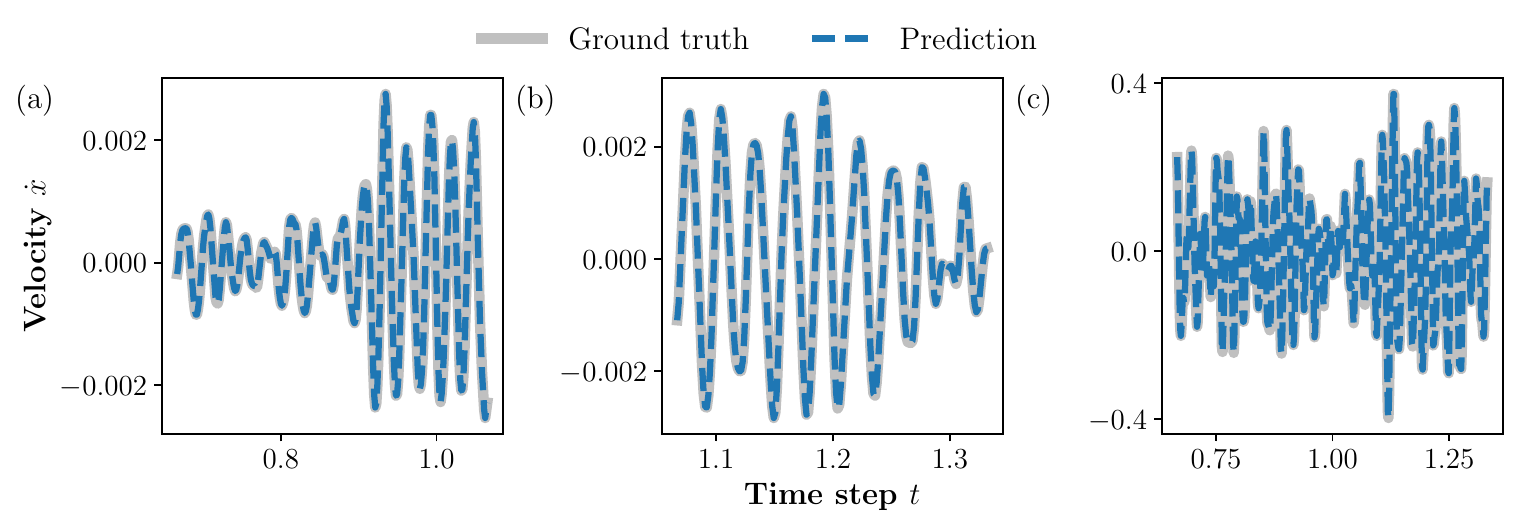}
    \caption{Velocity results on the benchmark system - without hysteresis. (a) Training results. (b) Testing 1 (same excitation) results. (c) Testing 2 (unseen excitation) results.}
    \label{fig:velocity results exp1 without hysteresis}
\end{figure}
\FloatBarrier

\end{revisedblock}

\clearpage
\bibliographystyle{elsarticle-num}   
\bibliography{refs}                  

\end{document}